\pdfoutput=1
\documentclass[11pt]{article}

\usepackage[final]{acl}

\usepackage{times}
\usepackage{latexsym}

\usepackage[T1]{fontenc}
\usepackage[utf8]{inputenc}

\usepackage{microtype}

\usepackage{inconsolata}

\usepackage{graphicx}
\usepackage{multirow}
\usepackage{booktabs}
\usepackage{amsmath}
\usepackage{amssymb}
\usepackage{makecell}
\usepackage{array}
\usepackage{enumitem}
\usepackage{float}
\setlist[itemize]{nosep, leftmargin=*}
\usepackage{tabularx}
\newcolumntype{C}{>{\centering\arraybackslash}X}

\title{CSAVocoder: A Causal Spatial Audio Vocoder\\Towards Real-Time Spatial Audio Generation}
\author{%
\normalsize
Zhiyuan Zhu\quad
Han Wang\quad
Wenxiang Guo\quad
Yu Zhang\quad \\
\textbf{Changhao Pan}\quad
\textbf{Rui Yang}\quad
\textbf{Zhou Zhao}\thanks{Corresponding Author}\quad\\
 Zhejiang University\\
\normalsize
\texttt{\{schmittzhu, zhaozhou\}@zju.edu.cn}\vspace{-1em}
}
\begin{document}

\maketitle

\begin{abstract}
Spatial audio vocoders are able to convert mel-spectrograms produced by generative models into spatial audio waveforms.
Most neural vocoders are designed for monaural audio, and direct extensions to spatial audio can degrade spatial quality by ignoring inter-channel cues.
We present CSAVocoder, a causal GAN-based spatial audio vocoder that jointly optimizes waveform fidelity and spatial rendering.
Our framework introduces a Spatial Adaptor that fuses multi-channel mel-spectrograms with dynamic source-listener pose information, together with a spatial consistency discriminator that supervises inter-channel cues.
To meet real-time requirements, we design a strictly causal, stateful generator that supports efficient streaming inference with constant memory overhead.   
Experiments on large-scale spatial audio datasets show that CSAVocoder improves spatial fidelity at competitive audio quality and real-time performance.
\end{abstract}

\section{Introduction}
\label{sec: intro}
Unlike monaural audio, spatial audio renders sound sources at different directions and distances, providing a more immersive listening experience.  
It reconstructs a three-dimensional sound field and exploits the natural localization mechanisms of the human auditory system.  
By accurately modeling these cues, spatial audio delivers a strong sense of presence and realism in digital environments.

Spatial audio is increasingly important in applications such as virtual reality, augmented reality~\cite{gupta2022augmented, kailas2021design}, and immersive gaming~\cite{raghuvanshi2018parametric, broderick2018importance, yadegari2022spatial}.  
Recent generative models have made progress in spatial audio synthesis~\cite{zhu2025asaudio, lu2025deep}, but many operate in the mel-spectrogram domain and rely on a vocoder to produce waveforms.
Works such as ISDrama~\cite{zhang2025isdrama} and DualSpec~\cite{zhao2026dualspec} use pretrained HiFi-GAN-style vocoders and achieve high single-channel quality, yet they largely ignore inter-channel spatial consistency.    
Most vocoder studies still target single-channel audio; direct extensions to spatial audio often degrade spatial quality because they ignore inter-channel cues.
The relative pose between the sound source and the listener is also a critical factor in spatial audio rendering.
% Accurate modeling of pose is essential for high-quality spatial audio.  
Recent works~\cite{immerse-diffusion-generative, 2024diffsage, templin2025sonicmotion} use various forms of spatial information, including explicit coordinates and features extracted from visual inputs.  
The relative position controls loudness and spectral coloration, while orientation affects perceived direction and spatial awareness.  
Therefore, an effective spatial audio vocoder must explicitly model and exploit pose information to improve both signal quality and spatial perception.

At the same time, real-time and efficiency requirements further complicate spatial audio rendering.
In virtual and augmented reality, user interaction and rapid scene changes require spatial audio to react with low latency in order to maintain immersion.  
Prior work~\cite{joy2024real, zhang2025isdrama} emphasizes real-time rendering and the real-time factor (RTF).  
Since the vocoder is the final stage of spatial audio generation, its inference speed directly impacts end-to-end system latency and is critical for real-time applications.

Designing a spatial audio vocoder that is both accurate and efficient is therefore challenging.
The model must synthesize waveforms with high fidelity.
It must also render perceptually valid spatial cues such as interaural level differences (ILD) and interaural phase differences (IPD), and learn the complex mapping from pose to acoustic behavior, including source position and motion.  
In addition, the vocoder must be causal and support low-latency streaming inference that generates audio continuously in chunks.

To address these challenges, we propose CSAVocoder, a causal GAN-based spatial vocoder that maps multi-channel mel-spectrograms and the relative source-listener pose to spatial waveforms and supports chunk-wise streaming inference.
In summary, our contributions are:
\begin{itemize}[nosep]
  \item We design a GAN-based spatial audio vocoder with a causal architecture that supports low-latency streaming inference while maintaining high-quality spatial audio synthesis.
  \item We introduce a pose-conditioning mechanism that uses a position adaptor to encode the relative source-listener pose and a mel adaptor to capture inter-channel relationships, improving spatial audio rendering and perceptual quality.
  \item We propose a unified architecture for the binaural and FOA formats considered in this work, learning an end-to-end mapping from multi-channel mel-spectrograms to multi-channel spatial audio waveforms.
\end{itemize}

\section{Related Work}
% https://github.com/vtuber-plan/hifi-gan.git: a implementation for 48kHz hifigan.
\label{sec: relatedwork}

\paragraph{Spatial Audio Rendering}
Spatial audio rendering constructs immersive auditory scenes by modeling sound propagation in three-dimensional space.  
Among existing representations, binaural audio and First-Order Ambisonics (FOA) are particularly central.  
Binaural audio directly models ear-canal signals via head-related transfer functions (HRTFs) and is the final perceptual format for headphone playback. 
FOA provides a spherical-harmonic, scene-centric representation with rotational equivariance and is widely used in VR/AR and 360$^\circ$ video systems.  
These two formats are therefore can be considered as the primary targets of many generative spatial audio models.

A broad line of work studies spatial audio generation and understanding from visual, textual, or multimodal inputs~\cite{zhu2026spatialomni}.  
2.5D Visual Sound~\cite{Gao_2019_CVPR} upmixes monophonic audio to binaural signals using visual cues in a regression setting.  
More recent methods move toward end-to-end spatial generation: ViSAGe~\cite{kim2025visage} predicts FOA from silent video, ISDrama~\cite{zhang2025isdrama} models long-form spatial narratives with explicit real-time constraints, Diff-SAGe~\cite{2024diffsage} applies diffusion in the complex spectral domain to better preserve inter-channel phase, and BEWO~\cite{sun2024both} enables text-driven binaural generation.  
ImmerseDiffusion~\cite{immerse-diffusion-generative}, In-the-Wild Audio Spatialization~\cite{pan-etal-2025-wild}, and streaming autoregressive diffusion models~\cite{lei2026swansphere} synthesize FOA or binaural audio from spatial and semantic conditions.  

Many of these systems operate primarily in the spectral domain and rely on separate vocoders or reconstruction stages, which introduce additional latency.  
Spatial information is often injected implicitly via latent variables or high-level prompts. Only a few works, such as ISDrama and ImmerseDiffusion, combine explicit spatial conditioning with considerations of real-time performance.  
These systems place strong demands on the spatial audio vocoder at the end of the pipeline: it must generate high-quality audio and preserve spatial cues, as measured by localization and listener-preference evaluations~\cite{pan2025spatialeval}.

\paragraph{Neural Vocoders}

Neural vocoders map acoustic features to waveforms and form the last stage of audio generation.  
GAN-based vocoders dominate due to favorable quality-efficiency trade-offs.  
HiFi-GAN~\cite{kong2020hifi} introduces multi-period and multi-scale discriminators; BigVGAN~\cite{lee2022bigvgan} improves robustness via periodic activations and anti-aliasing; FARGAN~\cite{fargan}, CARGAN~\cite{cargan}, and QGAN~\cite{chaudhary2024qgan} reduce parameters and computing complexity.
MusicHiFi~\cite{zhu2024musichifi} is an efficient high-fidelity stereophonic vocoder for enhancing low-resolution audio.

Alternative approaches operate in structured domains.  
Vocos~\cite{siuzdak2024vocosclosinggaptimedomain} predicts complex STFT coefficients; AF-Vocoder~\cite{chen2024afvocoder} applies frequency-domain artifact filtering; DisCoder~\cite{discoder} generates in the latent space of neural audio codecs.  
Diffusion vocoders such as DiffWave~\cite{kong2021diffwaveversatilediffusionmodel}, PriorGrad~\cite{lee2021priorgrad}, and FastDiff~\cite{huang2022fastdiff} offer high perceptual quality via iterative denoising, and flow-based vocoders such as WaveFM~\cite{luo2025wavefmhighfidelityefficientvocoder} learn direct transport trajectories for efficiency.
These existing vocoders primarily target monophonic or stereophonic audio and do not explicitly model spatial cues, limiting their effectiveness for spatial audio rendering.

\paragraph{Real-time Speech Synthesis}

Real-time synthesis is critical for interactive applications where latency must stay below perceptual thresholds, favoring causal architectures and streaming inference.  
Early neural vocoders such as WaveNet~\cite{van2016wavenet} are autoregressive and naturally causal, but sample-by-sample generation is too slow for real-time use.  
Online systems such as CONAN~\cite{zhang2025conanchunkwiseonlinenetwork} and Moshi~\cite{defossez2024moshi} use chunk-wise generation and state caching for bounded-delay interaction.  
For vocoders, WaveHax~\cite{wavehax} and MS-WaveHax~\cite{mswavehax2025} adopt causal convolutions with shuffle-based upsampling; DLL-APNet~\cite{du2025distilledlowlatencyneuralvocoder} combines distillation and simplification; MelFlow~\cite{welker2025realtimestreamingmelvocoding} adapts flow models to causal mel-to-waveform mapping; BinauralFlow~\cite{liang2025binauralflowcausalstreamableapproach} demonstrates streamable binaural generation.  
These advances motivate spatial vocoders that jointly achieve high spatial fidelity and streaming capability.

\section{Method}
\label{sec: method}
\subsection{Task Definition}
\label{sec: task_definition}
We aim to synthesize a multi-channel spatial audio waveform 
\(\mathbf{y} \in \mathbb{R}^{C \times L}\) 
from a multi-channel mel-spectrogram 
\(\mathbf{M} \in \mathbb{R}^{C \times F \times T}\) 
and the corresponding spatial pose sequence 
\(\mathbf{P} \in \mathbb{R}^{D_p \times T_p}\).
Here, \(C\) denotes the number of channels, \(L\) is the waveform length, \(F\) is the number of mel frequency bins, and \(T\) is the number of mel frames. 
The sequence \(\mathbf{P}\) captures the time-varying pose of the sound source relative to the listener, where \(D_p\) is the pose dimension and \(T_p\) is the number of pose samples. 
Each pose vector consists of a 3D Cartesian position \((x, y, z)\) and a 4D quaternion \((q_w, q_x, q_y, q_z)\) that encodes orientation, so \(D_p = 7\).

We formulate the problem as learning a conditional generative function \(G\) that maps the inputs to the target waveform:
\begin{equation}
  \mathbf{y} = G(\mathbf{M}, \mathbf{P}; \theta),
\end{equation}
where \(\theta\) denotes the learnable parameters of the generator.

\begin{figure*}[htbp] 
  \centering
  \includegraphics[
        % l, d, r, u
        trim={0cm 2.6cm 0cm 3.3cm}, 
        clip,
        width=0.9\textwidth]{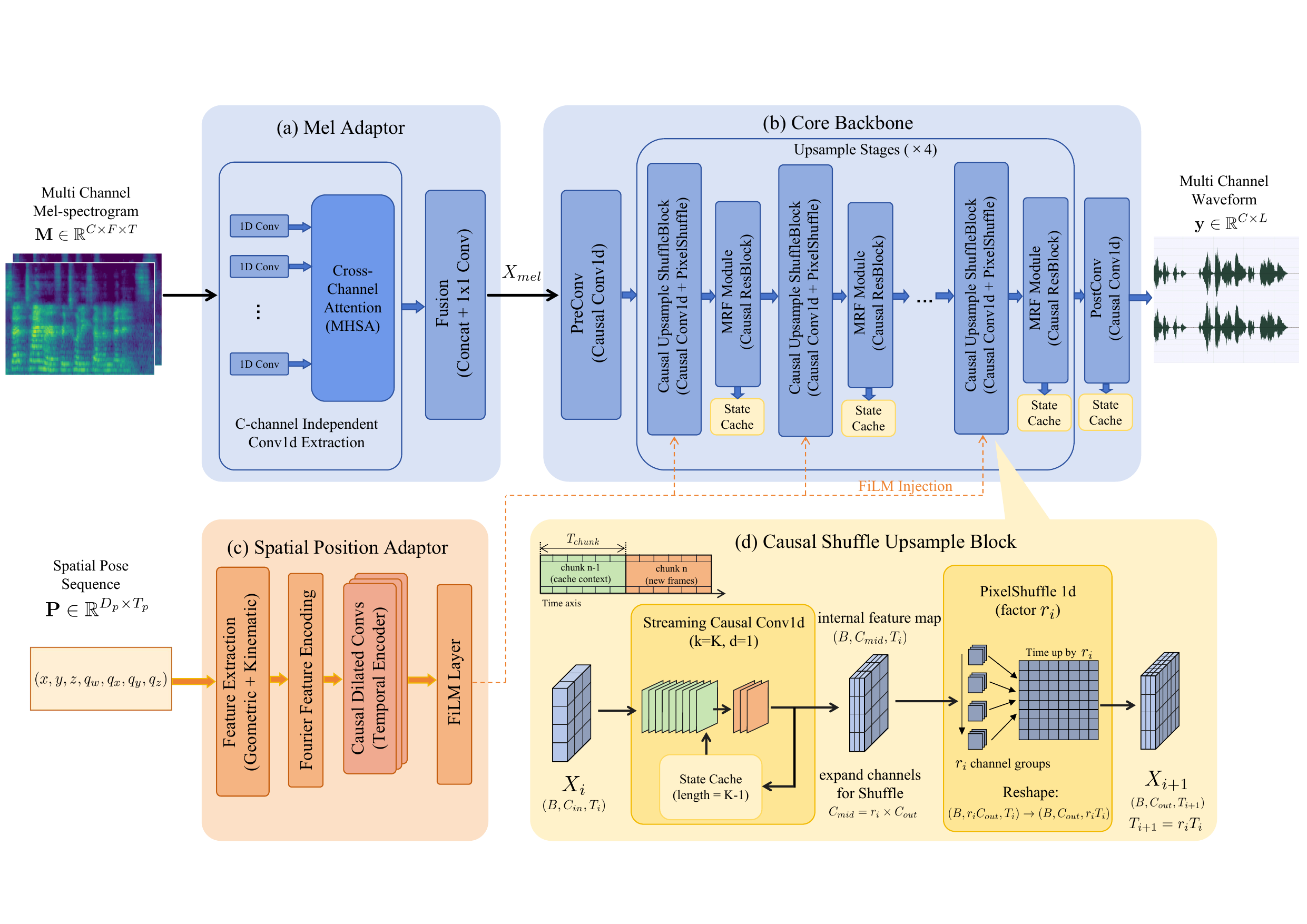}
  \caption{Overview of our model architecture.}
  \label{fig:model_arch}
\end{figure*}

\begin{figure*}[t]
  \centering
  \includegraphics[
        % l, d, r, u
        trim={0cm 1.7cm 0cm 1.35cm}, 
        clip, 
        width=0.9\textwidth]{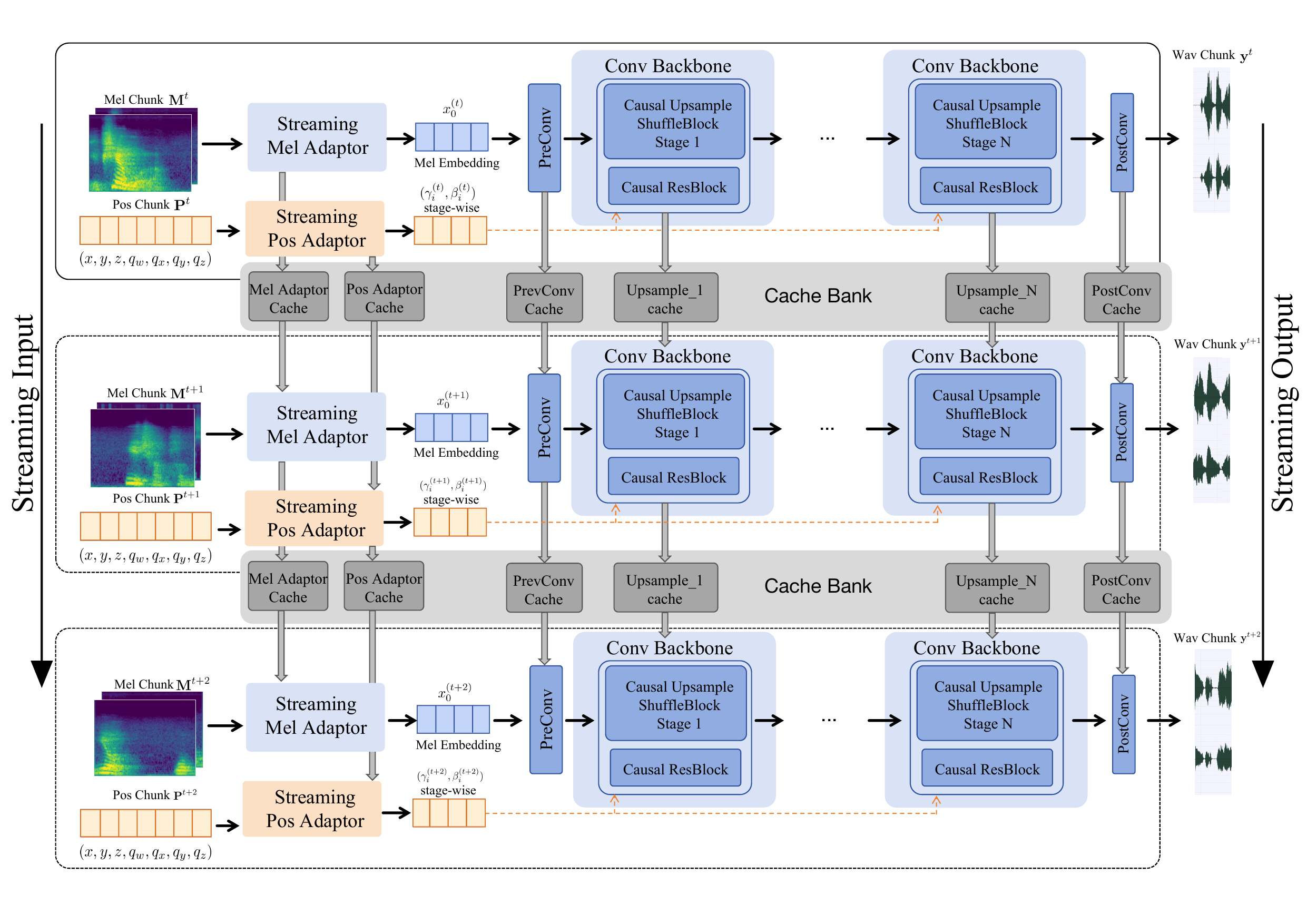}
  \caption{Continuous streaming inference pipeline. Starting from a multi-channel mel-spectrogram, the Mel Adaptor and Position Adaptor compute mel and pose embeddings. The embeddings are passed to the convolutional backbone, which generates waveform chunks through causal upsampling and residual blocks. Each streaming block keeps its own cache in the cache bank, so computation follows the causal constraint and never recomputes past activations.}
  \label{fig:streaming}
\end{figure*}

\subsection{GAN-based Vocoder}
\label{sec: gan_vocoder}
Our framework builds on the HiFi-GAN vocoder, which consists of a generator \(G\) and a set of discriminators \(D\). Figure~\ref{fig:model_arch} illustrates the overall architecture. We extend both the generator and the discriminator stack to support spatial conditioning and strictly causal streaming synthesis.

\subsubsection{Generator}
\label{sec: generator}
The generator follows the overall topology of HiFi-GAN, using a convolutional network to upsample the input mel-spectrogram in the temporal domain.
Instead of reusing the original non-causal modules, we redesign the core building blocks to satisfy strict causality and to support stateful streaming inference.

The generator takes the outputs of the Mel Adaptor and the Position Adaptor (Section~\ref{sec:spatial_adaptor}) as conditioning inputs.
The fused conditioning then passes through a series of upsampling and residual blocks, which gradually increase the temporal resolution to that of the target waveform.
We replace standard transposed convolutions with our ShuffleUpsampleBlock.
First, the {CausalConv1d} block projects the channels from \(C\) to \(C_{\text{out}} \cdot s\), producing
\(\mathbf{X}' \in \mathbb{R}^{B \times (C_{\text{out}} s) \times T_{\text{in}}}\).  
Then a ShuffleBlock reshapes this tensor to
\(\mathbf{X}'' \in \mathbb{R}^{B \times C_{\text{out}} \times (T_{\text{in}} s)}\) by folding extra channels into the time dimension.  
Since pixel shuffle is a pure tensor reordering without temporal mixing, it preserves the causality of the preceding convolution and yields artifact-free causal upsampling.

The residual blocks forming the multi-receptive-field fusion (MRF) stack are modified in the same spirit. 
Each {StreamingResBlock} consists of several causal convolutions with different dilation rates to capture patterns at multiple temporal scales, and maintains an internal buffer whose length matches its effective left context. 
These buffers store historical features from previous chunks, so even highly dilated convolutions receive the correct past context when processing the first frames of a new chunk.

\subsubsection{Discriminator}
\label{sec: discriminator}
\paragraph{Conventional Wave and Spectral Discriminators}
\label{sec: conventional_discriminators}
To encourage fidelity in both waveform and spectral domains, we combine the standard MPD and MSD from HiFi-GAN~\cite{kong2020hifi} with the MRD from BigVGAN~\cite{lee2022bigvgan}. 
The waveform discriminators capture periodic and multi-scale temporal artifacts, while the MRD uses multiple STFT configurations to detect frequency-domain artifacts at different resolutions.
When applied channel-wise or to flattened multi-channel signals, however, these discriminators are spatially agnostic: they cannot directly assess inter-channel relationships or the correctness of spatial cues.

\paragraph{Spatial Consistency Discriminator}
\label{sec: spatial_consistency_discriminator}
To explicitly supervise spatial structure, we introduce a Spatial Consistency Discriminator (SCD) that operates on multi-channel log-mel spectrograms and provides spatially informed adversarial gradients to the generator. 
Given a multi-channel waveform $\mathbf{y} \in \mathbb{R}^{B \times C \times T}$, the SCD computes $\mathbf{M} \in \mathbb{R}^{B \times C \times F \times T'}$ and projects it via a 2D convolution into latent features $\mathbf{X} \in \mathbb{R}^{B \times d \times C \times T'}$. 
An axial-attention backbone then applies multi-head self-attention (MHSA) along the temporal axis $(B \cdot C, T', d)$ and along the channel axis $(B \cdot T', C, d)$.
Temporal attention is bidirectional, since the discriminator processes finite segments offline.
Channel-axis attention models cross-channel spectral and level relationships.
A lightweight convolutional head maps the attended features to a scalar spatial-consistency score per segment, complementing conventional discriminators that mainly target single-channel fidelity.

\subsubsection{Training Objectives}
\label{sec: training_objectives}
To train the generator \(G\) and the discriminator set \(D\), we use a composite objective composed of several weighted loss terms.
We adopt the standard Least-Squares GAN adversarial loss ($\mathcal{L}_{\text{adv}}$) \cite{mao2017least}, feature matching loss ($\mathcal{L}_{\text{fm}}$) \cite{kumar2019melgangenerativeadversarialnetworks}, and multi-resolution spectral reconstruction losses ($\mathcal{L}_{\text{mel}}$ and $\mathcal{L}_{\text{STFT}}$) \cite{kong2020hifi}. The detailed formulations of these standard objectives are provided in Appendix~\ref{app: loss_design}.
Our primary contribution to the objective function is the format-aware Spatial Loss, designed to explicitly supervise spatial cues.

\paragraph{Spatial Loss}
\label{sec:spatial_loss}
Standard spectral losses treat channels independently and fail to constrain inter-channel spatial cues. 
We propose a format-aware spatial loss $\mathcal{L}_{\text{spatial}}$ that explicitly supervises physical attributes.

For binaural audio, following duplex theory, we combine interaural phase difference (IPD) and interaural level difference (ILD) losses: $\mathcal{L}_{\text{spatial}}^{\text{Bin}} = \lambda_{\text{IPD}} \mathcal{L}_{\text{IPD}} + \lambda_{\text{ILD}} \mathcal{L}_{\text{ILD}}$. 
The IPD loss operates on multi-resolution STFTs and compares phase differences in a sine-cosine embedding to avoid wrapping, with supervision concentrated in the low-frequency region by a Gaussian weighting. 
The ILD loss measures the discrepancy between the log-magnitude level differences of the two ears and emphasizes high frequencies through a complementary weighting.

For FOA audio, we define the loss on physical sound-field descriptors: $\mathcal{L}_{\text{spatial}}^{\text{FOA}} = \lambda_{\text{iv}} \mathcal{L}_{\text{iv\_dir}} + \lambda_{\text{r}} \mathcal{L}_{\text{r}} + \lambda_{\text{diff}} \mathcal{L}_{\text{diff}} + \lambda_{\text{elog}} \mathcal{L}_{\text{elog}}$. 
Direction-related terms ($\mathcal{L}_{\text{iv\_dir}}, \mathcal{L}_{\text{r}}$) constrain the angle and magnitude of the intensity vector with a low-frequency bias, while diffusion-related terms ($\mathcal{L}_{\text{diff}}, \mathcal{L}_{\text{elog}}$) capture ambient envelopment with a mid-high-frequency bias.

To stabilize training, all terms are modulated by an energy-based soft mask derived from the ground-truth signal. Detailed formulations are in Appendix~\ref{app: spatial_loss}.

\paragraph{Full Objective}
\label{sec: full_objective}
The total loss functions for the generator and the discriminators are defined as weighted sums of the components described above.

For each discriminator \(D_k\) in the discriminator set \(D\), the total loss consists only of the adversarial term:
\begin{align}
  \mathcal{L}_D = \sum_{k} \mathcal{L}_{\text{adv}}(D_k; G).
\end{align}

For the generator \(G\), the total loss is defined as
\begin{align}
  \mathcal{L}_G
  &= \mathcal{L}_{\text{adv}}(G; D)
  + \lambda_{\text{fm}} \mathcal{L}_{\text{fm}}
  + \lambda_{\text{mel}} \mathcal{L}_{\text{mel}} \notag \\
  &\quad + \lambda_{\text{STFT}} \mathcal{L}_{\text{STFT}}
  + \lambda_{\text{spatial}} \mathcal{L}_{\text{spatial}},
\end{align}
where \(\lambda_{\text{fm}}\), \(\lambda_{\text{mel}}\), \(\lambda_{\text{STFT}}\), and \(\lambda_{\text{spatial}}\) are hyperparameters that balance the contributions of different loss terms.

% This composite objective ensures that the model not only generates samples that are correct in a probabilistic sense, but also achieves high fidelity across multiple dimensions, including spectral structure, time-domain periodicity, and spatial perception.

\subsection{Causal Architecture for Streaming Synthesis}
\label{sec: causal_architecture}
Autoregressive vocoders such as WaveNet~\cite{van2016wavenet} are naturally causal but too slow due to sample-by-sample generation.
Non-autoregressive vocoders such as HiFi-GAN are fast, but their symmetric padding and transposed convolutions access future frames, which breaks temporal causality; simply switching to left-only padding often introduces boundary artifacts.

We therefore redesign the HiFi-GAN generator as a fully causal, explicitly stateful architecture tailored for streaming synthesis.  
All stages from mel features to waveform satisfy strict causality, and a stateful inference mechanism avoids redundant computation in chunk-based processing.  

\paragraph{Strict Causal Property}
When mapping a mel-spectrogram
\(
M = \{ m_1, \dots, m_T \}
\)
to a waveform
\(
W = \{ w_1, \dots, w_{T'} \}
\),
strict causality requires that each output sample \(w_t\) depends only on input frames
\(\{ m_1, \dots, m_i \}\) whose timestamps do not exceed that of \(w_t\).  
Any dependency on future frames \(m_j\) with \(j > i\) violates this constraint.  
Our design enforces this property at the operator level.
All convolutions use explicit left padding instead of symmetric padding.
For a kernel of size \(k\) and dilation \(d\), we pad \((k-1)d\) samples on the left and run the convolution with zero padding, so the output at any time step depends only on inputs at or before that step.

\iffalse
For completeness, we also implement two alternative fully causal upsampling modes:  
(i) a zero-insertion scheme followed by {CausalConv1d} , where zeros are inserted between samples and a causal filter learns interpolation;  
(ii) a padded transposed convolution that achieves causality via manual input padding and output trimming.  
\fi

\paragraph{Stateful Streaming Inference.}
Causality alone is insufficient for efficient streaming. Naively prepending a long contextual prefix to each chunk causes substantial redundant computation. 
We therefore implement all context-dependent layers in a stateful form. Each layer receives the current input chunk together with a compact cache from the previous step, and returns the current output plus an updated cache that stores exactly the left-context features required by the next chunk. 
As shown in Figure~\ref{fig:streaming}, the generator processes a sequence of mel chunks while propagating a global state object that aggregates the caches of all stateful layers, so no past activations are recomputed.
\subsection{Spatial Adaptor}
\label{sec:spatial_adaptor}
Standard mono-channel vocoders lack mechanisms to process multi-channel spectrograms or incorporate heterogeneous pose conditioning. 
We therefore introduce the Spatial Adaptor, which uses two parallel modules to encode spectral and geometric cues.

\paragraph{Attentional Mel Adaptor}
This module fuses the multi-channel mel-spectrogram $\mathbf{M} \in \mathbb{R}^{B \times C \times F \times T}$ into a unified single-stream representation $\mathbf{X}_{\text{mel}} \in \mathbb{R}^{B \times d_{\text{hifi}} \times T}$ while preserving implicit spatial cues.
First, we apply a shared weight-normalized 1D convolution to each channel independently to extract local features $\mathbf{X}_{\text{feat}} \in \mathbb{R}^{B \times C \times d \times T}$. 
To capture nonlinear inter-channel dependencies, we then employ Multi-Head Self-Attention along the channel axis at each time step. 
Unlike fixed difference operations, this data-driven approach dynamically weights the contribution of each channel. 
Finally, the attended features are concatenated and projected via a $1\times1$ convolution to the backbone dimension $d_{\text{hifi}}$, serving as the unified input to the generator.

\paragraph{Spatial Position Adaptor}
This adaptor converts the raw pose sequence $\mathbf{P}$ into dense, physically meaningful conditioning $\mathbf{X}_{\text{pos}}$. 
From the 7D raw pose, we derive Cartesian coordinates and forward vectors (from quaternions), augmented with first-order velocity differences to capture kinematic motion. 
To mitigate the spectral bias of MLPs, we map these scalars to high-dimensional sinusoidal representations using Fourier feature encoding~\cite{mildenhall2021nerf}, enabling sensitivity to fine-grained spatial changes.
The encoded features are then processed by {CausalPosEncoder}, a stack of causal dilated convolutions that models motion trajectories. 
We inject this condition into the generator via Feature-wise Linear Modulation (FiLM). 
For each upsampling block, audio features $\mathbf{x}_{\text{audio}}$ are modulated by scaling $\gamma$ and bias $\beta$ projected from the pose embeddings: $\text{FiLM}(\mathbf{x}_{\text{audio}}) = (1 + \tanh(\gamma)) \cdot \mathbf{x}_{\text{audio}} + \beta$.

\subsection{Unified Framework for Spatial Audio}
\label{sec: unified_framework}
Traditional vocoders are mono-centric or naïvely replicate single-channel outputs, limiting their applicability to spatial audio.
We design a channel-free generator where the shared backbone performs identical upsampling for any channel count. 
The Attentional Mel Adaptor fuses a \(C\)-channel mel-spectrogram into a fixed-dimensional representation, and the final projection layer outputs exactly \(C\) waveform channels.
For adversarial training, we pair this flexible generator with channel-aware discriminator heads specialized for each format.
The training objective is likewise format-aware: for 2-channel binaural samples the spatial loss emphasizes ILD and IPD, while for 4-channel FOA samples it switches to the sound-field losses above.
At inference, the model handles the supported formats by mapping the input mel-spectrogram and its channel configuration directly to spatial audio output.
The design keeps extension lightweight: supporting new standards such as 5.1 or 7.1 surround mainly requires format-specific objectives, discriminator heads, and evaluation protocols, while leaving the generator backbone unchanged.

\section{Experiments}
\label{sec: experiment}
\subsection{Experiment Details}
\paragraph{Dataset}
We use data in both binaural and FOA formats.
For binaural data, we adopt the MRSSpeech subset of MRSAudio~\cite{guo2025mrsaudio} and the EasyCom~\cite{donley2021easycom} dataset.
For FOA data, we use the Spatial LibriSpeech~\cite{sarabia2023spatial} dataset, synthesized from LibriSpeech~\cite{panayotov2015librispeech}, which offers a large number of FOA samples with spatial annotations. 
To increase spatial and acoustic diversity, we further generate simulated data using the SoundSpaces toolkit.
In total, our training corpus contains roughly 600 hours of binaural data (about 350k samples) and 900 hours of FOA data (about 310k samples), all stored as 16-bit PCM at a sampling rate of 48 kHz.

We preprocess EasyCom and MRSSpeech datasets using the ClearVoice~\cite{zhao2025clearervoice} denoising algorithm to enhance audio quality.
We randomly sample 700 segments from all datasets as the test set, and split the remaining data into training and validation sets with a 9:1 ratio.
% We also use binaural dataset from \cite{leng2022binauralgrad} as an unseen test set.
Detailed statistics are provided in Appendix~\ref{app:dataset}.

\begin{table*}[t]
\centering
\small
\setlength{\tabcolsep}{2pt}
\renewcommand{\arraystretch}{1.0}
\begin{tabularx}{\textwidth}{l|CC|CCCCCCC}
\toprule
\textbf{Model} &
% \textbf{ANG COS $\uparrow$} &
\makecell{\textbf{ANG }\\\textbf{COS} ($\uparrow$)} &
% \textbf{DIS COS $\uparrow$} &
\makecell{\textbf{DIS }\\\textbf{COS} ($\uparrow$)} &
% \textbf{l2 $\downarrow$} &
% \mbox{\textbf{Amplitude} $\downarrow$} &
\textbf{MRSTFT ($\downarrow$)} &
% \textbf{Phase $\downarrow$} &
\textbf{PESQ ($\uparrow$)} &
\textbf{MCD ($\downarrow$)} &
\mbox{\textbf{Periodicity($\downarrow$)}} &
% \textbf{V/UV F1 ($\uparrow$)} &
\textbf{RTF ($\downarrow$)} \\
\midrule
HiFi-GAN        & 39.07             & 68.37            & 1.470             & 1.562             & 5.329             & 0.169             & \underline{0.0622} \\
CARGAN          & 30.00             & 63.71            & 1.194             & 1.739             & 3.377             & 0.160             & 0.1348 \\
FARGAN          & 23.53             & 56.03            & 1.219             & 1.885             & 3.447             & 0.161             & 0.1916 \\
DiffWave        & 27.05             & 61.07            & 1.585             & 2.070             & 6.440             & 0.156             & 0.1621 \\
PriorGrad       & 37.18             & 64.71            & 1.981             & 2.364             & 3.671             & 0.128             & 0.1629 \\
FastDiff        & 29.46             & 59.93            & 2.889             & 2.051             & 6.948             & 0.130             & 0.1174 \\
WaveFM          & \underline{41.36} & \underline{71.96}& \underline{1.079} & \underline{2.400} & 2.727             & 0.141             & 0.1634 \\
Vocos           & 40.04             & 70.23            & \textbf{1.039}    & \textbf{2.510}    & \textbf{1.892}    & \underline{0.113} & \textbf{0.0339} \\
\midrule
\textbf{Ours}   & \textbf{62.11}   & \textbf{77.05}   & {1.223}    & 2.109             & \underline{2.153}    & \textbf{0.107}    & {0.1587} \\
\bottomrule
\end{tabularx}
\caption{Objective comparison with vocoder baselines on the binaural test set (spatial metrics in percentages).}
\label{tab:metrics}
\end{table*}

\begin{table*}[t]
\centering
\footnotesize
\setlength{\tabcolsep}{5pt}
\renewcommand{\arraystretch}{1.0}
\begin{tabular}{>{\scriptsize}l|ccccccc}
\toprule
\textbf{Method} & \textbf{ANG COS} ($\uparrow$) & \textbf{DIS COS} ($\uparrow$) & \textbf{MRSTFT} ($\downarrow$) & \textbf{PESQ} ($\uparrow$) & \textbf{MCD} ($\downarrow$) & \textbf{Per.} ($\downarrow$) & \textbf{RTF} (s1+s2) \\
\midrule
{GT + DSP}                 & 26.65 & 51.07 & \textbf{1.189} & \textbf{3.884} & \underline{4.353} & \underline{0.090} & 0 + 0.2030 \\
Vocos + DSP              & 23.65 & 50.02 & 1.347 & 2.216 & 6.126 & 0.103 & 0.0170 + 0.2030 \\
GT + BinauralGrad        & \underline{50.83} & \underline{72.13} & 2.536 & \underline{3.548} & 8.815 & \textbf{0.078} & 0 + 1.1282 \\
Vocos + BinauralGrad     & 46.45 & 69.59 & 2.692 & 2.149 & 8.948 & 0.113 & 0.0170 + 1.1282 \\
\midrule
ISDrama + HiFi-GAN       & 45.77 & 67.70 & 2.782 & 1.110 & 8.885 & 0.206 & 0.183 + 0.062 \\
ISDrama + Ours           & 47.73 & 71.44 & 2.676 & 1.090 & 9.454 & 0.219 & 0.183 + 0.1587 \\
\midrule
\textbf{GT + Ours}       & \textbf{62.11} & \textbf{77.05} & \underline{1.223} & 2.109 & \textbf{2.153} & 0.107 & 0 + 0.1587 \\
\bottomrule
\end{tabular}
\caption{Comparison with DSP and two-stage spatialization baselines (spatial metrics in percentages).}
\label{tab:two_stage_baselines}
\end{table*}

\paragraph{Baseline}
We compare our proposed method with several vocoder baselines.
We use the original HiFi-GAN~\cite{kong2020hifi}, Vocos~\cite{siuzdak2024vocosclosinggaptimedomain}, CARGAN~\cite{cargan}, FARGAN~\cite{fargan}, and WaveFM~\cite{luo2025wavefmhighfidelityefficientvocoder} as baselines.
We further include diffusion vocoders DiffWave~\cite{kong2021diffwaveversatilediffusionmodel}, PriorGrad~\cite{lee2021priorgrad}, and FastDiff~\cite{huang2022fastdiff}.
Recent work such as MusicHiFi~\cite{zhu2024musichifi} has explored stereophonic vocoding, but its implementation is not publicly available. Because dedicated spatial vocoder baselines are limited, we use the above models, which have shown strong performance in monaural audio generation.
We perform channel-wise inference on all baselines to generate binaural and FOA audio for comparison with our model.
DSP and two-stage spatialization baselines are compared in Section~\ref{sec:spatialization_baselines}.

\paragraph{Metrics}
Our evaluation protocol comprises both subjective listening tests and objective metrics. 

The objective evaluation covers general audio quality, spectral and temporal similarity, and spatial characteristics. 

For waveform and spectral similarity, we adopt metrics used in BinauralGrad~\cite{leng2022binauralgrad}:
% Wave L2 , 
% Amplitude L2 and Phase L2 computed on STFT representations to quantify magnitude and phase discrepancies, 
MCD (mel-cepstral distortion) to measure spectral distortion,
periodicity to assess harmonic consistency,
and MRSTFT, which combines spectral convergence with log- and linear-magnitude terms to improve spectral alignment. 
We also report PESQ as a perceptual measure for speech-related quality assessment. 
Except for PESQ, lower metric values indicate better performance.

To quantify spatial fidelity, we introduce two consistency measures, ANG Cos and DIS Cos, which respectively evaluate angular and distance similarity between generated and reference signals. 
Specifically, we extract angular and distance embeddings from binaural audio using Spatial-AST. 
Because Spatial-AST~\cite{zheng2024bat} produces position estimates only for static sources, 
we partition each audio into 1-second segments, 
compute the cosine similarity between predicted and ground-truth embeddings within each segment, 
and then average these segment-level similarities to obtain an overall spatial-consistency score.
These metrics are reported as percentages.

For subjective evaluation, we use MOS-Q (Mean Opinion Score for Quality) to evaluate the quality of generated
audio and MOS-P (Mean Opinion Score for Position) to assess spatial perception. More implementation
details are in Appendix~\ref{app:experiment_details}.

\subsection{Quantitative Comparison}

We compare our model with existing vocoder baselines and present the metric results in Table~\ref{tab:metrics}. 
As shown in the table, our approach improves the spatial metrics over all baselines, including the diffusion vocoders, while remaining competitive on audio metrics.
These gains indicate that explicitly modeling inter-channel relationships through our Mel Adaptor and supervising spatial cues via the Spatial Consistency Discriminator are important for preserving spatial information.
The main audio-quality trade-off appears in PESQ.
Our PESQ score is lower than non-causal baselines such as Vocos and WaveFM. This is partly expected under the strictly causal constraint, since our causal convolutions only access past context, whereas non-causal models can use bidirectional receptive fields that benefit perceptual quality.
More results on FOA are in Appendix~\ref{app: foa_res}.
Analyses on bandwidth-matched evaluation, causal variants, and pose perturbations are in Appendix~\ref{app:additional_binaural}.

We report the Real-Time Factor (RTF) measured on a single NVIDIA RTX 4090 GPU.
Our model achieves RTF = 0.1587, which is well below unity and indicates that the causal streaming architecture supports real-time generation. 
Compared to other GAN-based baselines, our RTF remains competitive while providing substantially better spatial preservation.
Detailed latency results are provided in Appendix~\ref{sec:appendix-latency}.

Overall, these results show that CSAVocoder achieves both high-fidelity waveform synthesis and accurate spatial rendering.
The causal architecture introduces a minor quality trade-off compared to non-causal models, but this is an acceptable cost for enabling low-latency streaming applications.

\subsection{Comparison with Spatialization Baselines}
\label{sec:spatialization_baselines}
Channel-wise vocoder baselines share our mel-to-waveform input condition, so we further compare against spatialization pipelines with different input assumptions.
The DSP baseline computes propagation delays from source-ear geometry and applies channel-wise time warping to a mono waveform.
BinauralGrad~\cite{leng2022binauralgrad} serves as a learned waveform spatializer, and ISDrama~\cite{zhang2025isdrama} provides an upstream spatial generation stage that produces binaural mel-spectrograms.
Each pipeline combines a stage-1 input (GT mono source, Vocos output, or ISDrama mel) with a stage-2 rendering module (DSP, BinauralGrad, HiFi-GAN, or CSAVocoder), and we report the RTF of both stages.
GT in stage 1 denotes ground-truth mono audio and GT in last row denotes binaural mel-spectrogram from GT binaural audio.

Table~\ref{tab:two_stage_baselines} shows that GT mel-spectrogram with our vocoder reaches 62.11/77.05 ANG/DIS COS, higher than DSP method with GT wav input (26.65/51.07) and BinauralGrad with GT wav input (50.83/72.13).
Replacing the stage-1 input with Vocos output degrades all spatialization pipelines, which confirms that upstream mono quality bounds the final spatial fidelity.
ISDrama + Ours also outperforms ISDrama + HiFi-GAN (47.73/71.44 vs. 45.77/67.70), indicating that our vocoder preserves spatial cues even under upstream generated mel inputs.
This comparison clarifies the positioning of our work: the main contribution is a spatial vocoder under matched mel-to-waveform constraints, whereas traditional rendering systems assume different inputs such as mono sources, HRTFs, and separate spatialization modules.

\subsection{Qualitative Comparison}
\begin{figure*}[htbp] 
  \centering
  \includegraphics[width=\textwidth]{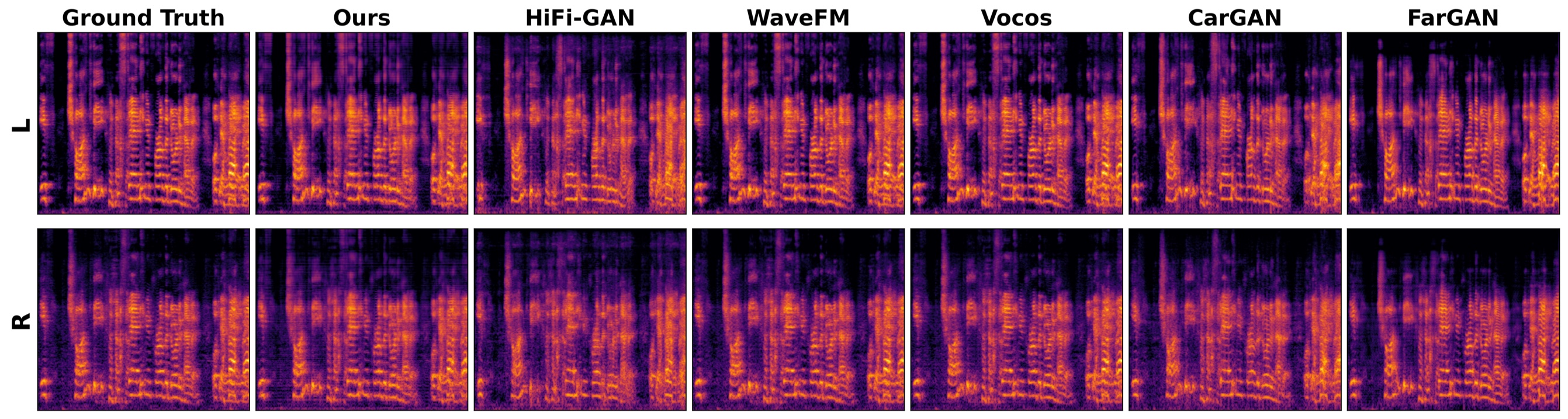}
  \caption{Qualitative comparison of spectrograms. The first column shows the ground truth, the second column shows our model, and the remaining columns show baselines.}
  \label{fig:qualitative_comparison}
\end{figure*}

We conduct a qualitative comparison of our proposed model with the baselines.
Figure~\ref{fig:qualitative_comparison} compares generated audio samples.
The first column shows the ground-truth audio, the second column shows audio generated by our model, and the remaining columns show baseline predictions.
Our causal model preserves harmonic stacks and formant trajectories, which closely match the ground truth on both channels, while maintaining consistent left-right spectral patterns. 
Compared with the baselines, our results exhibit sharper and more coherent harmonic structures with fewer band-wise artifacts and a cleaner noise floor. 
Our causal generation still shows slightly smoother transients and mildly reduced high-frequency detail than non-causal counterparts. Nevertheless, it achieves a highly similar overall spectral structure, indicating that high perceptual quality is attainable under causal constraints.

\subsection{Subjective Evaluation}
We conduct subjective listening tests to evaluate the quality and spatial perception of the generated audio.

Table~\ref{tab:subjective_evaluation} reports the subjective evaluation results.
For the MOS-P test, listeners rate how accurately they perceive the sound-source position in the generated audio compared to the ground-truth position on a scale from 1 to 5.
Our model achieves the highest MOS-P score among all models, indicating superior spatial perception.
For the MOS-Q test, listeners rate the overall audio quality of the generated samples on a scale from 1 to 5.
Our model also achieves a high MOS-Q score, although causal generation can be slightly weaker in audio quality than non-causal models that use future context.
We further conduct a MUSHRA-style expert evaluation on spatial perception, reported in Appendix~\ref{app:mushra}.

\begin{table}[H]
\centering
\begin{tabular}{lcc}
\toprule
\textbf{Model} &
\textbf{MOS-P} & 
\textbf{MOS-Q} \\
\midrule
% mean+-std
HiFi-GAN & $3.86 \pm 0.19$ & $3.98 \pm 0.17$ \\
CARGAN   & $3.90 \pm 0.18$ & $4.03 \pm 0.14$ \\
FARGAN   & $3.93 \pm 0.14$ & $4.07 \pm 0.15$ \\
WaveFM   & $4.13 \pm 0.13$ & $4.17 \pm 0.12$ \\
Vocos    & $4.09 \pm 0.15$ & $4.24 \pm 0.11$ \\
\midrule
\textbf{Ours} & ${4.25} \pm 0.16$ & ${4.09} \pm 0.21$\\
\midrule
GT & $4.42 \pm 0.11$ & $4.41 \pm 0.16$ \\
\bottomrule
\end{tabular}

\caption{MOS results with 95\% confidence intervals.}
\label{tab:subjective_evaluation}
\end{table}

\subsection{Ablation Study}

We perform ablation studies of our proposed components and present the results in Table~\ref{tab:ablation_study}. 

All proposed components contribute to the overall performance of our model.
Removing the Mel Adaptor causes a clear drop in spatial metrics, since the model no longer uses inter-channel information.
Using 4 attention heads in the Mel Adaptor yields the best performance.
Removing the Position Adaptor results in a moderate degradation, indicating that spatial information can still be partially captured through the Mel Adaptor.
The results also show that the Spatial Consistency Discriminator improves spatial metrics. In our experiments, however, its adversarial weight required careful tuning; otherwise, training could become unstable.
\begin{table}[ht]
\centering
\begin{tabular}{l|cc}
\toprule
\textbf{Setting} &
\makecell{\textbf{ANG}\\\textbf{COS} ($\uparrow$)} &
\makecell{\textbf{DIS}\\\textbf{COS} ($\uparrow$)}  \\
\midrule
w/o Mel Adaptor& 42.60 & 65.39 \\
Mel Adaptor 2 head & 61.03 & 76.55\\
Mel Adaptor 8 head & 61.50 & 76.70\\
w/o SCD &  58.82 & 74.63\\
w/o Position Adaptor & 54.78 & 70.63 \\
\midrule
Mel Adaptor 4 head & \textbf{62.11} & \textbf{77.05}\\
\bottomrule
\end{tabular}
\caption{Ablation of adaptor components and attention head counts on the binaural test set.}
\label{tab:ablation_study}
\end{table}

\section{Conclusion}
\label{sec: con}

We present CSAVocoder, a spatial audio vocoder that jointly addresses high-fidelity waveform synthesis and accurate spatial rendering.  
Our framework extends the GAN architecture with three key innovations: a Spatial Adaptor that fuses multi-channel mel-spectrograms with dynamic pose information to capture inter-channel relationships, a Spatial Consistency Discriminator that explicitly supervises spatial cues, and a strictly causal, stateful generator whose streaming inference incurs only constant memory overhead.

Experimental results show that CSAVocoder outperforms channel-wise vocoder baselines in spatial fidelity while maintaining strong audio quality and real-time performance.
Within the binaural and FOA settings studied here, the unified architecture supports multiple spatial formats in a single framework, making it a practical solution for immersive audio applications such as virtual reality, augmented reality, and spatial communication.

These results highlight explicit spatial modeling and causal streaming as a strong foundation for future work on real-time spatial audio generation.

\clearpage

\section*{Limitations}
Our work has three main limitations.
First, establishing fair comparisons against causal baselines is challenging because different implementations adopt distinct buffering strategies and runtime optimizations that affect both latency and quality.
Many strong vocoders are optimized for offline generation and benefit from non-causal context or heavier post-processing; even when adapted to streaming, their engineering choices can dominate measured runtime.
A standardized causal-baseline suite with matched end-to-end latency budgets and consistent objective measurements is left for future work.
Second, we focus on binaural and FOA formats; extending to higher-order ambisonics (HOA), multichannel loudspeaker layouts (e.g., 5.1/7.1), object-based audio, and personalized HRTF rendering is non-trivial.
Increasing channel counts changes the required inductive bias, stability of adversarial training, and computational cost, and different ambisonic conventions may introduce dataset mismatches.
Third, we condition on pose (position and orientation), but alternative or complementary representations may be more robust or expressive.
Examples include relative geometry features (distance/azimuth/elevation), scene-aware embeddings from visual or 3D context, or learned spatial tokens that summarize multi-source environments.
We leave these axes to future work.

\section*{Ethical Considerations}
This paper presents CSAVocoder, a causal and stateful vocoder for low-latency spatial audio generation conditioned on acoustic features. 
While the model does not generate linguistic content on its own, it can be integrated into upstream TTS/VC systems; therefore, both model- and data-related risks must be considered.

\paragraph{Data provenance, licensing, and privacy.}
We rely on publicly available speech/spatial-audio corpora and simulation pipelines. We do not claim ownership of any third-party audio content and recommend that any release avoid redistributing raw audio unless explicitly permitted by original licenses/terms. Derived artifacts such as file lists, splits, and evaluation scripts should be shared in a way that enables reproducibility while reducing privacy exposure. Speech datasets may contain personally identifying information or sensitive attributes.

\paragraph{Risks from real-time generation and speech privacy.}
Low-latency speech generation can enable near-real-time impersonation, “live” spoofing in voice authentication, and the re-synthesis of intercepted private conversations. Spatial audio further increases realism and may strengthen deceptive scenarios. In addition, pose conditioning introduces an auxiliary privacy surface: logged 3D trajectories and orientations can reveal behavioral patterns, attention, or activity context in immersive systems.

\paragraph{Potential harmful applications.}
Beyond deepfakes, potential misuse includes covert surveillance, harassment, social engineering, or generating misleading evidence. Dataset misuse may include training downstream models for speaker identification, demographic profiling, or other applications that participants did not consent to, especially when data is repurposed outside its original scope.

\paragraph{Mitigations and responsible release.}
We recommend (i) clear acceptable-use terms and licenses; (ii) optional watermarking/provenance signals and guidance for detection; (iii) restricting and documenting deployment contexts; (iv) minimizing retention of raw audio, intermediate representations, and pose logs; and (v) reporting limitations and failure modes. For listening tests, risks are minimal but include fatigue; conservative volume, breaks, and withdrawal options are advised.

\paragraph{Bias and environmental impact.}
Training data and simulators may under-represent languages, accents, acoustic environments, and accessibility-related speech characteristics, leading to uneven performance. Finally, while causal inference can reduce runtime cost, training remains compute-intensive; we encourage transparent reporting of compute and settings to support reproducibility and responsible scaling.

\bibliography{custom}

\newpage
\appendix
\section{Implementation Details}
\label{app: implementation_details}

This appendix provides the hyperparameters and architectural configurations used in our experiments.
\subsection{Audio and Spectrogram Parameters}

All audio processing and mel-spectrogram extraction are conducted using the parameters listed in Table~\ref{tab:audio_params}.  
The generator upsampling factor is set to 320 to match the mel-spectrogram hop size.

\begin{table}[h]
  \centering
  \caption{Audio processing and mel-spectrogram extraction parameters}
  \label{tab:audio_params}
  \begin{tabular}{lc}
    \toprule
    \textbf{Parameter} & \textbf{Value} \\
    \midrule
    Sample rate & 48,000 Hz \\
    FFT size & 1024 \\
    Hop size & 320 \\
    Window size & 1024 \\
    Number of mel bins & 128 \\
    Mel $f_{\min}$ & 20 Hz \\
    Mel $f_{\max}$ & 24,000 Hz \\
    \bottomrule
  \end{tabular}
\end{table}

\subsection{Generator Architecture}

The generator backbone $G$ is based on the HiFi-GAN V1 configuration and is modified to support causal streaming synthesis.  
The total upsampling factor is \(8 \times 5 \times 4 \times 2 = 320\).  
The detailed configuration is provided in Table~\ref{tab:generator_arch}.
\begin{table*}[ht]
\centering
\small
\setlength{\tabcolsep}{4pt}
\renewcommand{\arraystretch}{1.2}
\begin{tabularx}{\textwidth}{lCCCC}
    \toprule
    \textbf{Layer / Block} & \textbf{Output Channels} & \textbf{Kernel} & \textbf{Stride} & \textbf{Upsample} \\
    \midrule
    Initial conv (\texttt{conv\_pre}) & 512 & 7 & 1 & -- \\
    \midrule
    Upsampling block 1 & & & & \\
    \quad Causal upsampling & 256 & 16 & 8 & $\times 8$ \\
    \quad MRF residual blocks & 256 & [3, 7, 11] & -- & -- \\
    \midrule
    Upsampling block 2 & & & & \\
    \quad Causal upsampling & 128 & 10 & 5 & $\times 5$ \\
    \quad MRF residual blocks & 128 & [3, 7, 11] & -- & -- \\
    \midrule
    Upsampling block 3 & & & & \\
    \quad Causal upsampling & 64 & 8 & 4 & $\times 4$ \\
    \quad MRF residual blocks & 64 & [3, 7, 11] & -- & -- \\
    \midrule
    Upsampling block 4 & & & & \\
    \quad Causal upsampling & 32 & 4 & 2 & $\times 2$ \\
    \quad MRF residual blocks & 32 & [3, 7, 11] & -- & -- \\
    \midrule
    Final conv (\texttt{conv\_post}) & $C$ & 7 & 1 & -- \\
    \bottomrule
  \end{tabularx}
\caption{Generator backbone configuration}
\label{tab:generator_arch}
\end{table*}

\subsection{Spatial Adaptor Architecture}

The Spatial Adaptor consists of two core submodules: the Attentional Mel Adaptor and the Spatial Position Adaptor.  
Their configurations are summarized in Table~\ref{tab:adaptor_arch}.

\begin{table*}[ht]
  \centering
  \setlength{\tabcolsep}{20pt}
  \begin{tabularx}{\textwidth}{llc}
    \toprule
    \textbf{Submodule} & \textbf{Hyperparameter} & \textbf{Value} \\
    \midrule
    \multirow{4}{*}{Attentional Mel Adaptor} 
      & Input mel bins & 128 \\
      & Hidden channels & 256 \\
      & Conv kernel size & 5 \\
      & Number of attention heads & 4 \\
    \midrule
    \multirow{6}{*}{Spatial Position Adaptor} 
      & Input pose dimension & 7 \\
      & Fourier feature bands & 8 \\
      & Causal temporal encoder layers & 3 \\
      & Temporal encoder kernel size & 3 \\
      & Injection mechanism & FiLM \\
      & Injection feature dimension & 256 \\
    \bottomrule
  \end{tabularx}
  \caption{Spatial adaptor configuration}
  \label{tab:adaptor_arch}
\end{table*}

\subsection{Training and Optimization Hyperparameters}

The training and optimization hyperparameters are listed in Table~\ref{tab:training_params}.  
We adopt a standard adversarial training setup with additional spectral and spatial losses.

\begin{table}[t]
  \centering
  \setlength{\tabcolsep}{4pt}
  \begin{tabular}{lc}
    \toprule
    \textbf{Hyperparameter} & \textbf{Value} \\
    \midrule
    Optimizer & Adam \\
    Learning rate (G / D) & $2 \times 10^{-4}$ \\
    Adam betas $(\beta_1, \beta_2)$ & (0.8, 0.99) \\
    Learning rate decay $\gamma$ & 0.999 \\
    Batch size & 16 \\
    Audio segment length & 16,384 samples \\
    \midrule
    \multicolumn{2}{c}{\textbf{Loss weights}} \\
    \midrule
    $\mathcal{L}_{\text{adv}}$ ($\lambda_{\text{adv}}$) & 1.0 \\
    $\mathcal{L}_{\text{fm}}$ ($\lambda_{\text{fm}}$) & 2.0 \\
    $\mathcal{L}_{\text{mel}}$ ($\lambda_{\text{mel}}$) & 45.0 \\
    $\mathcal{L}_{\text{STFT}}$ ($\lambda_{\text{STFT}}$) & 1.0 \\
    $\mathcal{L}_{\text{spatial}}$ (IPD/ILD) & 0.1 \\
    $\mathcal{L}_{\text{spatial}}$ (FOA) & 2.0 \\
    \bottomrule
  \end{tabular}
  \caption{Training and optimization hyperparameters}
  \label{tab:training_params}
\end{table}

\section{Losses Design}
\label{app: loss_design}

\subsection{Adversarial Objective ($\mathcal{L}_{\text{adv}}$)}
\label{app: adversarial_objective}
We adopt the Least-Squares GAN (LS-GAN) for adversarial training.  
For each discriminator \(D_k\) in the set \(\{D_k\}\), the discriminator loss is
\(
  \mathcal{L}_{\text{adv}}(D_k, G)
  = \mathbb{E}_{\mathbf{y}}\big[(D_k(\mathbf{y}) - 1)^2\big]
  + \mathbb{E}_{\mathbf{z}, \mathbf{c}}\big[D_k(G(\mathbf{z}, \mathbf{c}))^2\big],
\)
which encourages high scores for real samples \(\mathbf{y}\) and low scores for generated samples \(G(\mathbf{z}, \mathbf{c})\).

The generator adversarial loss is
\(
  \mathcal{L}_{\text{adv}}(G, D)
  = \sum_{k} \mathbb{E}_{\mathbf{z}, \mathbf{c}}
    \big[(D_k(G(\mathbf{z}, \mathbf{c})) - 1)^2\big],
\)
which encourages all discriminators to regard generated audio as real.

Since \(D\) comprises the MPD, MSD, MRD, and SCD introduced above, the final adversarial objectives are
\(
  \mathcal{L}_{\text{adv}}(G) = \sum_{k} \mathcal{L}_{\text{adv}}(G; D_k),
  \mathcal{L}_{\text{adv}}(D) = \sum_{k} \mathcal{L}_{\text{adv}}(D_k; G),
\)
which jointly enforce alignment with real audio in temporal structure, multi-scale patterns, spectral detail, and spatial consistency.

\subsection{Feature Matching Loss ($\mathcal{L}_{\text{fm}}$)}
\label{app: feature_matching_loss}
To stabilize GAN training and regularize the generator toward the real data manifold, we employ a feature matching loss.  
It acts as a perceptual constraint based on learned hierarchical representations:
\(
  \mathcal{L}_{\text{fm}}(G, D)
  = \sum_{k} \mathbb{E}_{\mathbf{y}, \mathbf{z}, \mathbf{c}}
    \left[
      \sum_{i=1}^{L_k}
      \frac{1}{N_i}
      \big\|
        D_k^{(i)}(\mathbf{y}) - D_k^{(i)}(G(\mathbf{z}, \mathbf{c}))
      \big\|_1
    \right],
\)
where \(D_k^{(i)}\) is the \(i\)-th intermediate feature map of discriminator \(D_k\), \(L_k\) is the number of layers considered, and \(N_i\) is the number of elements in that feature map.

\subsection{Auxiliary Perceptual and Reconstruction Losses}
\label{app: auxiliary_losses}
These losses provide more direct, non-adversarial gradient signals to the generator and optimize specific perceptual aspects of the synthesized audio.

To ensure that the spectral structure of the generated audio matches that of real audio, we employ two spectral reconstruction losses.

The first is the mel-spectrogram loss \(\mathcal{L}_{\text{mel}}\), which computes the L1 distance between the mel-spectrograms of the generated audio \(G(\mathbf{M}, \mathbf{P})\) and the real audio \(\mathbf{y}\).  
This loss constrains the model on the perceptually important mel scale and is defined as
\begin{equation}
  \mathcal{L}_{\text{mel}}(G)
  = \mathbb{E}_{\mathbf{y}, \mathbf{M}, \mathbf{P}}
    \big[ \big\| \phi(\mathbf{y}) - \phi(G(\mathbf{M}, \mathbf{P})) \big\|_1 \big],
\end{equation}
where \(\phi\) denotes the transformation from the waveform to its mel-spectrogram.

The second is the multi-resolution STFT loss \(\mathcal{L}_{\text{STFT}}\).  
This loss is computed under multiple short-time Fourier transform (STFT) configurations, each with different FFT sizes, window sizes, and hop sizes.  
It consists of two components: the spectral convergence loss \(\mathcal{L}_{\text{sc}}\), which penalizes differences in spectral magnitude, and the log STFT magnitude loss \(\mathcal{L}_{\text{mag}}\), which computes an L1 loss on the log-magnitude spectrogram and better reflects human perception of loudness.  
The total STFT loss is defined as the average of these two components across all STFT resolutions.

\subsection{Spatial Loss Formulation}
\label{app: spatial_loss}
We provide the full formulation of the spatial loss \(\mathcal{L}_{\text{spatial}}\), which explicitly supervises inter-channel spatial cues beyond per-channel spectral similarity.  
Its concrete form is defined in a format-adaptive way for binaural and First-Order Ambisonics (FOA) signals.

\paragraph{Binaural Spatial Loss.}
For binaural signals, we compute complex STFTs of the left and right channels, \(S_L(f,t)\) and \(S_R(f,t)\), under multiple STFT configurations.  
The interaural phase difference (IPD) is given by \(\Delta \Phi(f,t) = \arg S_L(f,t) - \arg S_R(f,t)\).  
To avoid phase wrapping, we embed \(\Delta \Phi\) into the complex plane and define
\[
  \mathbf{u}_{\text{IPD}}(f,t)
  = \bigl(\cos \Delta \Phi(f,t),\, \sin \Delta \Phi(f,t)\bigr) \in \mathbb{R}^2.
\]
The IPD loss compares the embedded representations of the target and generated signals,
\[
  \resizebox{\columnwidth}{!}{$
  \mathcal{L}_{\text{IPD}}
  = \frac{\sum_{f,t} w_{\text{IPD}}(f) \, m(f,t)
      \, \bigl\|\mathbf{u}_{\text{IPD}}^{\text{pred}}(f,t)
               - \mathbf{u}_{\text{IPD}}^{\text{ref}}(f,t)\bigr\|_2^2}
         {\sum_{f,t} w_{\text{IPD}}(f) \, m(f,t) + \varepsilon},$}
\]
where \(w_{\text{IPD}}(f) = \exp\bigl(-(f / f_{\text{IPD,max}})^2\bigr)\) emphasizes low frequencies and \(m(f,t)\) is an energy-based soft mask.  

The interaural level difference (ILD) is defined in the log-magnitude domain as
\[
  \resizebox{\columnwidth}{!}{$
  \text{ILD}^{\text{ref}}(f,t)
  = 20 \log_{10} |S_L^{\text{ref}}(f,t)| -
    20 \log_{10} |S_R^{\text{ref}}(f,t)|,$}
\]
and analogously for \(\text{ILD}^{\text{pred}}\).  
The ILD loss is
\[
  \resizebox{\columnwidth}{!}{$
  \mathcal{L}_{\text{ILD}}
  = \frac{\sum_{f,t} w_{\text{ILD}}(f) \, m(f,t)
      \,\bigl|\text{ILD}^{\text{pred}}(f,t)
               - \text{ILD}^{\text{ref}}(f,t)\bigr|}
         {\sum_{f,t} w_{\text{ILD}}(f) \, m(f,t) + \varepsilon},$}
\]
with \(w_{\text{ILD}}(f) = 1 - \exp\bigl(-(f / f_{\text{ILD,min}})^2\bigr)\) that emphasizes high frequencies.  

The soft mask \(m(f,t)\) is derived from the frame-wise energy of the reference signal.  
Let \(E(t)\) be the RMS energy at frame \(t\) (averaged over frequency and channels), and
\[
  E_{\mathrm{dB}}(t) = 10 \log_{10}(E(t) + \varepsilon).
\]
We define a smooth frame-wise speech activity
\[
  s(t) = \sigma\!\left(\frac{E_{\mathrm{dB}}(t) - \mu_{\text{VAD}}}{\sigma_{\text{VAD}}}\right),
\]
where \(\sigma(\cdot)\) is the sigmoid function, \(\mu_{\text{VAD}}\) is the soft-VAD center in dB, and \(\sigma_{\text{VAD}}\) controls the transition width.  
The time–frequency mask is then
\[
  m(f,t) = m_{\min} + (1 - m_{\min})\, s(t),
\]
with \(m_{\min} > 0\) to avoid nullifying silent regions.  
The binaural spatial loss is
\[
  \mathcal{L}_{\text{spatial}}^{\text{bin}}
  = \lambda_{\text{IPD}} \mathcal{L}_{\text{IPD}}
  + \lambda_{\text{ILD}} \mathcal{L}_{\text{ILD}}.
\]

\paragraph{FOA Spatial Loss.}
For FOA signals, we assume a B-format ordering \((W, X, Y, Z)\). 
Given target and predicted waveforms \(y, \hat{y} \in \mathbb{R}^{B \times 4 \times T}\), we compute complex STFTs for each scale, obtaining
\[
  W(f,t),\, X(f,t),\, Y(f,t),\, Z(f,t)
\]
for the reference and \(\hat{W}(f,t),\, \hat{X}(f,t),\, \hat{Y}(f,t),\, \hat{Z}(f,t)\) for the prediction. 
The total FOA energy at each bin is
\begin{align*}
  E^{\text{ref}}(f,t) &= |W|^2 + |X|^2 + |Y|^2 + |Z|^2, \\
  E^{\text{pred}}(f,t) &= |\hat{W}|^2 + |\hat{X}|^2 + |\hat{Y}|^2 + |\hat{Z}|^2.
\end{align*}

\textit{Energy-weighted mask and frequency biases.}
We reuse the soft mask \(m(f,t)\) from the binaural case, now interpreted per FOA STFT configuration. 
To steer supervision across frequency, we define a low-frequency bias for direction-related terms,
\[
  w_{\text{dir}}(f) = \exp\bigl(-(f / f_{\text{iv,max}})^2\bigr),
\]
and a smooth mid–high-frequency bias for diffuseness-related terms. 
Let \(f_s\) be the sampling rate and \(\tilde{f} = f / (f_s/2)\) the normalized frequency. 
We set
\[
  w_{\text{diff}}(f)
  = \tfrac{1}{2} + \tfrac{1}{2}
    \tanh\!\left(\frac{\tilde{f} - c_{\text{diff}}}{w_{\text{diff}}}\right),
\]
where \(c_{\text{diff}}\) controls the center of the transition and \(w_{\text{diff}}\) controls its width. 

We also apply mild energy exponents \(E^{\alpha}\) to emphasize high-energy bins without dominating the loss. 
We denote these exponents by \(\alpha_{\text{iv}}, \alpha_{\text{r}}, \alpha_{\text{diff}}\).

\textit{Intensity vector and directional term.}
The active intensity components are computed as
\begin{align*}
  I_X^{\text{ref}}(f,t) &= \Re\{W^*(f,t) X(f,t)\}, \\
  I_Y^{\text{ref}}(f,t) &= \Re\{W^*(f,t) Y(f,t)\}, \\
  I_Z^{\text{ref}}(f,t) &= \Re\{W^*(f,t) Z(f,t)\},
\end{align*}
and analogously for \(I_X^{\text{pred}}, I_Y^{\text{pred}}, I_Z^{\text{pred}}\). 
We collect these into intensity vectors
\begin{align*}
  \mathbf{I}^{\text{ref}}(f,t) &= [I_X^{\text{ref}}, I_Y^{\text{ref}}, I_Z^{\text{ref}}]^\top, \\
  \mathbf{I}^{\text{pred}}(f,t) &= [I_X^{\text{pred}}, I_Y^{\text{pred}}, I_Z^{\text{pred}}]^\top.
\end{align*}
The directional mismatch is measured via the cosine distance
\[
  d_{\text{iv}}(f,t)
  = 1 - \frac{
      \mathbf{I}^{\text{ref}}(f,t)^\top
      \mathbf{I}^{\text{pred}}(f,t)
    }{
      \|\mathbf{I}^{\text{ref}}(f,t)\|_2
      \,\|\mathbf{I}^{\text{pred}}(f,t)\|_2 + \varepsilon
    },
\]
and we define
\[
  \resizebox{\columnwidth}{!}{$
  \mathcal{L}_{\text{iv\_dir}}
  = \frac{\sum_{f,t} m(f,t)\, w_{\text{dir}}(f)\,
           \bigl(E^{\text{ref}}(f,t)\bigr)^{\alpha_{\text{iv}}}
           \, d_{\text{iv}}(f,t)}
         {\sum_{f,t} m(f,t)\, w_{\text{dir}}(f)\,
           \bigl(E^{\text{ref}}(f,t)\bigr)^{\alpha_{\text{iv}}}
          + \varepsilon}.$}
\]

\textit{Normalized intensity ratio term.}
We normalize the intensity by total energy,
\begin{align*}
  \mathbf{r}^{\text{ref}}(f,t) &= \frac{\mathbf{I}^{\text{ref}}(f,t)}{E^{\text{ref}}(f,t) + \varepsilon}, \\
  \mathbf{r}^{\text{pred}}(f,t) &= \frac{\mathbf{I}^{\text{pred}}(f,t)}{E^{\text{pred}}(f,t) + \varepsilon},
\end{align*}
and define
\[
  \resizebox{\columnwidth}{!}{$
  \mathcal{L}_{\text{r}}
  = \frac{\sum_{f,t} m(f,t)\, w_{\text{dir}}(f)\,
           \bigl(E^{\text{ref}}(f,t)\bigr)^{\alpha_{\text{r}}}
           \,\bigl\|\mathbf{r}^{\text{pred}}(f,t)
                    - \mathbf{r}^{\text{ref}}(f,t)\bigr\|_1}
         {\sum_{f,t} m(f,t)\, w_{\text{dir}}(f)\,
           \bigl(E^{\text{ref}}(f,t)\bigr)^{\alpha_{\text{r}}}
          + \varepsilon}.$}
\]

\textit{Diffuseness term.}
We compute the intensity norm
\[
  \|\mathbf{I}^{\text{ref}}(f,t)\|_2,\quad
  \|\mathbf{I}^{\text{pred}}(f,t)\|_2,
\]
and define diffuseness as
\begin{align*}
  D^{\text{ref}}(f,t) &= 1 - \frac{\|\mathbf{I}^{\text{ref}}(f,t)\|_2}{E^{\text{ref}}(f,t) + \varepsilon}, \\
  D^{\text{pred}}(f,t) &= 1 - \frac{\|\mathbf{I}^{\text{pred}}(f,t)\|_2}{E^{\text{pred}}(f,t) + \varepsilon}.
\end{align*}
The diffuseness loss is then
\[
  \resizebox{\columnwidth}{!}{$
  \mathcal{L}_{\text{diff}}
  = \frac{\sum_{f,t} m(f,t)\, w_{\text{diff}}(f)\,
           \bigl(E^{\text{ref}}(f,t)\bigr)^{\alpha_{\text{diff}}}
           \,\bigl(D^{\text{pred}}(f,t)
                  - D^{\text{ref}}(f,t)\bigr)^2}
         {\sum_{f,t} m(f,t)\, w_{\text{diff}}(f)\,
           \bigl(E^{\text{ref}}(f,t)\bigr)^{\alpha_{\text{diff}}}
          + \varepsilon}.$
    }
\]

\textit{Log-energy term.}
Finally, we align the log-energy fields of reference and prediction:
\begin{align*}
  \log E^{\text{ref}}(f,t) &= \log(E^{\text{ref}}(f,t) + \varepsilon), \\
  \log E^{\text{pred}}(f,t) &= \log(E^{\text{pred}}(f,t) + \varepsilon),
\end{align*}

and define
\[
  \resizebox{\columnwidth}{!}{
  $\displaystyle
    \mathcal{L}_{\text{elog}}
    = \frac{\sum_{f,t} m(f,t)\, w_{\text{diff}}(f)\,
             \bigl|\log E^{\text{pred}}(f,t)
                   - \log E^{\text{ref}}(f,t)\bigr|}
           {\sum_{f,t} m(f,t)\, w_{\text{diff}}(f) + \varepsilon}
  $}%
\]

\textit{Multi-scale aggregation.}
In practice, all the above quantities are computed for multiple STFT parameter sets \((n_{\text{FFT}}, \text{hop}, \text{win})\). 
The four FOA terms \(\mathcal{L}_{\text{iv\_dir}}, \mathcal{L}_{\text{r}}, \mathcal{L}_{\text{diff}}, \mathcal{L}_{\text{elog}}\) are averaged over scales, and the final FOA spatial loss is
\[
  \mathcal{L}_{\text{spatial}}^{\text{FOA}}
  = \lambda_{\text{iv}} \mathcal{L}_{\text{iv\_dir}}
  + \lambda_{\text{r}} \mathcal{L}_{\text{r}}
  + \lambda_{\text{diff}} \mathcal{L}_{\text{diff}}
  + \lambda_{\text{elog}} \mathcal{L}_{\text{elog}}.
\]

\section{Details of Datasets}
\label{app:dataset}
\subsection{Recorded Binaural and FOA Data}
We use both binaural and First-Order Ambisonics (FOA) spatial audio data for training and evaluation. 
For the binaural branch, we adopt the MRSSpeech subset of the MRSAudio~\cite{guo2025mrsaudio} corpus together with the EasyCom~\cite{donley2021easycom} dataset, which contain extensive indoor recordings captured with binaural microphones. 
These corpora cover multiple speakers, diverse source-listener spatial configurations, and both Chinese and English speech, providing realistic binaural characteristics and room acoustics. 
For FOA, we use the Spatial LibriSpeech~\cite{sarabia2023spatial} dataset, which is synthesized from LibriSpeech~\cite{panayotov2015librispeech} and provides a large number of FOA-format spatial speech samples with corresponding position annotations. 
However, Spatial LibriSpeech only models azimuthal variation on the horizontal plane during spatialization and lacks diversity along the vertical dimension (elevation).
As a result, many samples contain little variation in the $z$ channel, which can make the model overfit horizontal-plane cues and reduce its sensitivity to vertical direction.
Assembling recorded corpora at this scale requires controlled capture and dense manual labeling.
Similar data-construction efforts appear in other audio areas, including spatial and singing corpus construction, transcription, and long-form benchmarking~\cite{guo2025stars, zhang2024gtsinger, li2024robust, pan2026swanbench}.

\subsection{Simulated Spatial Data from SoundSpaces (MP3D)}
To enrich spatial diversity, especially in elevation and in complex 3D room geometries, we additionally generate a large amount of simulated spatial data based on the SoundSpaces~\cite{NEURIPS2022_3a48b0ea} simulation framework and Habitat-Sim~\cite{Savva_2019_ICCV}. 
In this work we focus on indoor scenes from the Matterport3D (MP3D) dataset; for each MP3D environment we instantiate a Habitat-Sim simulator and attach an audio sensor configured either as binaural (2-channel) or FOA Ambisonics (4-channel) at a sampling rate of 48\,kHz, following procedure in~\cite{liu2026jaeger}. 
The listener (receiver) is placed at a height of 1.5\,m above the floor, and the audio materials configuration from MP3D is loaded to enable frequency-dependent reflection, absorption, and diffraction in the propagation engine.
We calculate the relative pose between source and receiver, and use it as conditioning input to the model.

\subsection{Static BRIR/RIR Sampling and Position Generation.}
For the static subset, we randomly sample receiver and source positions on the MP3D navigation mesh. 
A candidate pair is accepted only if the horizontal distance lies within $(1,10)$\,m and the height difference is smaller than 2\,m, which avoids degenerate configurations (too close or too far, or across floors). 
For each accepted pair we query the audio sensor once and obtain a binaural or FOA room impulse response (BRIR/RIR). 
All positions are initially given in the Habitat/BAT coordinate convention, where the horizontal plane is $x$-$z$, $y$ points upwards, and the agent faces the $-z$ direction. 
For downstream usage we convert all 3D positions $(x,y,z)$ into a more conventional, listener-centric coordinate system with the horizontal plane being $x$-$y$, $z$ pointing upwards, and the listener facing the $+y$ direction.
All relative positions (source minus receiver) in our dataset use this coordinate system. 

\subsection{Dynamic Simulated Trajectories.}
Besides the purely static BRIRs, we also construct a dynamic subset in which the listener remains fixed while the source moves through the environment. 
Concretely, for a given receiver position we randomly sample two source points that are both within a reasonable distance from the receiver and compute the shortest path between them on the navigation mesh. 
The resulting 3D path is uniformly subsampled to a fixed number of time steps (e.g., 20 frames per trajectory). 
At each step we update the source position in Habitat-Sim, query a new BRIR from the audio sensor, and record the corresponding source position, relative position, and coarse direction labels (\texttt{left/right}, \texttt{front/behind}, \texttt{above/below}) derived from the transformed coordinate system. 
For each utterance we also generate a frame-level pose sequence at 20\,Hz by repeating the (static) relative position or by aligning it with the dynamic trajectory, yielding an $N\times7$ pose matrix per audio sample that is fully time-synchronized with the waveform.

\subsection{Convolution with Mono Speech and Post-processing.}
To turn the simulated BRIR/RIRs into training data, we convolve them with clean, single-channel speech from the LibriSpeech~\cite{panayotov2015librispeech} corpus. 
All LibriSpeech utterances are first resampled to 48 kHz and converted to mono.
For each utterance we randomly select one BRIR entry, perform FFT-based convolution to obtain either 2-channel binaural or 4-channel FOA audio, and then truncate the result to match the original utterance length. 
We apply simple peak normalization (with a conservative safety margin) to avoid clipping and ensure that all simulated samples are loudness-consistent with the real-world data. 
Pairing simulation with recording lets us cover configurations that are hard to collect directly.
Comparable strategies, combining large-scale data or pretrained generative priors, are widely adopted in other audio generation tasks~\cite{feng2026rectifying, pan2026audioediting}.

\subsection{Overall Dataset Scale}
Combining the real and simulated corpora, our final training and evaluation set comprises approximately 600 hours of binaural data and 900 hours of FOA data. 
Among them, around 220k binaural samples and 70k FOA samples are synthesized by convolving LibriSpeech with SoundSpaces-generated BRIR/RIRs in MP3D environments, while the remaining samples come from MRSSpeech, EasyCom, and Spatial LibriSpeech. 
All audio is uniformly resampled to 48 kHz, and all spatial annotations are provided in the unified listener-centric coordinate system.
This scale is comparable to the annotated corpora that expressive singing synthesis systems rely on for controllable style and technique~\cite{zhang2024tcsinger, zhang2025tcsinger, zhang2024stylesinger, guo2025techsinger}.

\section{Additional Binaural Evaluation}
\label{app:additional_binaural}

We provide additional binaural analyses on sampling-rate fairness, causality, and pose robustness.

\subsection{Bandwidth-matched Evaluation}
To control for native sampling-rate differences across baselines, we additionally evaluate all methods under a common effective bandwidth. 
We apply a low-pass filter with cutoff \(f_c=7.8\) kHz to both references and outputs before resampling to the rate required by each metric.

\begin{table*}[htbp]
\centering
\small
\setlength{\tabcolsep}{4pt}
\renewcommand{\arraystretch}{1.0}
\begin{tabularx}{\textwidth}{lCCCCCC}
\toprule
\textbf{Model} &
\makecell{\textbf{ANG}\\\textbf{COS} ($\uparrow$)} &
\makecell{\textbf{DIS}\\\textbf{COS} ($\uparrow$)} &
\makecell{\textbf{MRSTFT}\\($\downarrow$)} &
\makecell{\textbf{PESQ}\\($\uparrow$)} &
\makecell{\textbf{MCD}\\($\downarrow$)} &
\makecell{\textbf{Per.}\\($\downarrow$)} \\
\midrule
HiFi-GAN          & 50.24 & 75.02 & 2.0317 & 1.3397 & 9.8823 & 0.1824 \\
HiFi-GAN (48kHz) & 64.77 & 80.48 & 0.9999 & 2.1226 & 3.4236 & 0.1305 \\
CARGAN            & 57.21 & 77.89 & 1.1352 & 1.7441 & 3.9013 & 0.1114 \\
FARGAN            & 42.00 & 67.84 & 1.0677 & 1.8937 & 4.1077 & 0.1154 \\
WaveFM            & 66.05 & 82.84 & \textbf{0.8593} & \textbf{2.6007} & 2.7301 & \textbf{0.0916} \\
Vocos             & 68.64 & 83.46 & 0.8840 & 2.5119 & \textbf{2.3722} & 0.0931 \\
\textbf{Ours}     & \textbf{68.71} & \textbf{83.86} & 1.1440 & 2.0752 & 2.5072 & 0.0933 \\
\bottomrule
\end{tabularx}
\caption{Bandwidth-matched evaluation with a common low-pass cutoff \(f_c=7.8\) kHz.}
\label{tab:bandwidth_matched}
\end{table*}

As shown in Table~\ref{tab:bandwidth_matched}, the relative ranking on audio metrics remains similar, and our method retains leading spatial consistency after bandwidth matching.
The smaller spatial gap compared with the full-band setting is expected because high-frequency cues are important for binaural spatial perception.

\subsection{Causal and Non-causal Variants}
We compare causal and non-causal variants to isolate the effect of strict streaming constraints.

\begin{table*}[htbp]
\centering
\small
\scriptsize
\setlength{\tabcolsep}{1pt}
\begin{tabularx}{\textwidth}{lCCCCCC}
\toprule
\textbf{Model} & \makecell{\textbf{ANG}\\\textbf{COS}} & \makecell{\textbf{DIS}\\\textbf{COS}} & \textbf{MRSTFT} & \textbf{PESQ} & \textbf{MCD} & \textbf{Per.} \\
\midrule
Vocos & 40.04 & 70.23 & \textbf{1.039} & \textbf{2.510} & 1.892 & 0.113 \\
Causal Vocos & 41.06 & 71.35 & 1.240 & 2.240 & 2.977 & 0.112 \\
\makecell[l]{Non-causal CSAVocoder} & 60.21 & 74.29 & 1.083 & 2.397 & \textbf{1.742} & \textbf{0.093} \\
\makecell[l]{\textbf{Causal} \textbf{CSAVocoder}} & \textbf{62.11} & \textbf{77.05} & 1.223 & 2.109 & 2.153 & 0.107 \\
\bottomrule
\end{tabularx}
\caption{Causal and non-causal comparison. ANG/DIS COS and PESQ are higher-is-better; others are lower-is-better.}
\label{tab:causal_variants}
\end{table*}

Table~\ref{tab:causal_variants} indicates that non-causal context improves several audio-quality metrics, but both CSAVocoder variants substantially outperform Vocos variants on spatial metrics. 
This pattern indicates that the spatial gains mainly come from spatial conditioning and inter-channel modeling rather than from causality alone.

\subsection{Pose Perturbation Robustness}
We evaluate robustness to noisy or partially missing pose inputs by adding Gaussian noise or randomly dropping pose entries during inference. 
The average pose standard deviation per audio sample is approximately 1.1, so the tested Gaussian noise levels correspond to moderate perturbations.

\begin{table}[htbp]
\centering
\small
\begin{tabular}{llcc}
\toprule
\textbf{Noise Type} & \textbf{Level} & \makecell{\textbf{ANG}\\\textbf{COS} ($\uparrow$)} & \makecell{\textbf{DIS}\\\textbf{COS} ($\uparrow$)} \\
\midrule
Gaussian noise & 0.05 & 62.10 & 77.01 \\
Gaussian noise & 0.1  & 61.34 & 76.00 \\
Gaussian noise & 0.2  & 59.11 & 74.83 \\
Random drop    & 10\% & 61.19 & 77.04 \\
Random drop    & 20\% & 61.67 & 76.11 \\
Random drop    & 30\% & 60.18 & 75.82 \\
w/o Pose       & --   & 56.61 & 75.75 \\
\textbf{CSAVocoder} & -- & \textbf{62.11} & \textbf{77.05} \\
\bottomrule
\end{tabular}
\caption{Robustness to pose noise and missing entries.}
\label{tab:pose_robustness}
\end{table}

Table~\ref{tab:pose_robustness} shows gradual degradation under stronger perturbations, while the model stays robust to moderate noise and pose dropout.

\section{FOA Results}
\label{app: foa_res}
\begin{table*}[htbp]
\centering
\small
\setlength{\tabcolsep}{2pt}
\renewcommand{\arraystretch}{1.0}
\begin{tabularx}{\textwidth}{l|CC|CCCCCCC}
\toprule
\textbf{Model} &
\textbf{Corr\_all ($\uparrow$)} &
\textbf{AUC\_ j \_all ($\uparrow$)} &
\textbf{MRSTFT ($\downarrow$)} &
\textbf{MCD (dB) ($\downarrow$)} &
\textbf{PESQ ($\uparrow$)} \\
\midrule
HiFi-GAN        & 18.65         & 61.98            & 1.278             & 4.052             & 2.122             \\
CARGAN          & 15.99         & 61.20            & 1.257             & 3.690             & 1.757             \\
FARGAN          & 16.26         & 61.28            & 1.154             & 2.941             & 1.794             \\
WaveFM          & 14.37         & 60.80            & 0.846             & 1.950             & 3.520             \\
Vocos           & 19.49         & 62.92            & 0.918             & 1.453             & 2.997             \\
\midrule
\textbf{Ours}   & {18.53}       & {63.44}          & {1.248}           & 3.449             & {1.972}           \\
\bottomrule
\end{tabularx}
\caption{FOA Results}
\label{tab:foa_metrics}
\end{table*}
We present additional experimental results for FOA spatial audio synthesis. 
For FOA audio, we adopt similar evaluation metrics as for binaural audio, including audio quality metrics (PESQ, MRSTFT, MCD) and spatial consistency metrics (Corr\_all and AUC\_j\_all).
The spatial consistency metrics are derived from the ViSAGe work~\cite{kim2025visage} and assess the ability of the generated audio to preserve spatial cues.
For audio quality, we use common metrics such as PESQ, MRSTFT, and MCD to measure the quality of the generated audio.
For the FOA format, we specifically evaluate the audio quality of the \(W\) channel.
% Additionally, we report the model's Real-Time Factor (RTF) to assess inference speed.
Table~\ref{tab:foa_metrics} presents a quantitative comparison of our method against several baseline models on the FOA spatial audio synthesis task. 
The results indicate competitive FOA spatial preservation, with the best AUC\_j\_all among the compared models.
Its audio-quality metrics remain comparable to those of non-causal models.

\section{Latency Evaluation}
\label{sec:appendix-latency}

This appendix defines the latency measures used for streaming inference and reports representative results under different chunk sizes.

\subsection{Definitions}
For streaming audio generation, we consider three types of latency:
\paragraph{Algorithmic latency ($L_{\text{alg}}$, ms).}
Algorithmic latency is the inherent delay introduced by the streaming design, independent of hardware speed. Under chunked inference, a system that outputs audio only after receiving a full chunk has a lower bound
\begin{equation}
L_{\text{alg}} \ge T_{\text{chunk}} + T_{\text{lookahead}} + T_{\text{overlap}},
\end{equation}
where $T_{\text{chunk}}$ is the chunk duration, $T_{\text{lookahead}}$ is any future-context requirement (0 for strictly causal designs), and $T_{\text{overlap}}$ accounts for cross-fade/overlap-add schemes that require waiting for future samples.
For our model, $T_{\text{lookahead}}=0$ and $T_{\text{overlap}}=0$.

\paragraph{Compute latency ($L_{\text{comp}}$, ms/chunk).}
Compute latency is the wall-clock time required to run the model for one chunk (forward pass in streaming mode). We report distributional statistics (p50/p90/p99) because tail latency is critical for real-time playback stability.

\paragraph{Real-Time Factor (RTF).}
To compare compute latency across chunk sizes, we use
\begin{equation}
\text{RTF} = \frac{L_{\text{comp}}}{T_{\text{chunk}}}.
\end{equation}
RTF $< 1$ indicates faster-than-real-time inference. 

\subsection{Chunking under \texorpdfstring{$\text{sr}=48$ kHz, hop=320}{sr=48 kHz, hop=320}}
With sampling rate $\text{sr}=48$ kHz and hop size 320 samples, the feature frame rate is
\begin{equation}
f = \frac{48000}{320} = 150~\text{frames/s},
\end{equation}
Thus, chunk sizes of 40/60/80/100 ms correspond to 6/9/12/15 mel frames, respectively.

\subsection{Measurement protocol}
We benchmark streaming inference with batch size 1 and disable gradient computation. 
For GPU timing, we synchronize before and after each forward pass to measure true kernel execution time. 
We perform a warm-up phase to avoid one-time compilation and cache effects, then run a fixed number of iterations and collect per-chunk latency samples, from which we compute mean and percentiles (p50/p90/p99). 
Unless stated otherwise, $L_{\text{comp}}$ includes only model inference and excludes feature extraction, file I/O, and audio device buffering.

\subsection{Results and discussion}
Table~\ref{tab:latency} reports representative compute latency under different chunk sizes. Across repeated runs, the mean compute latency stays in a narrow band (approximately 15 ms/chunk), while RTF improves as chunk size increases.
This behavior is expected on GPUs when sequence lengths are short: fixed overheads (kernel launches, framework scheduling, memory movements) can dominate, and larger chunks may better utilize the GPU, reducing the \emph{per-frame} cost even if ms/chunk is similar. 
All tested settings achieve RTF $< 1$, indicating real-time feasibility with substantial headroom.

\begin{table}[t]
\centering
\small
\begin{tabular}{c c c c c c}
\toprule
Chunk & Mean & p50 & p90 & p99 & RTF \\
\midrule
40  & $15.24 \pm 0.95$  & 14.99 & 16.44 & 18.62 & 0.3811 \\
60  & $15.15 \pm 1.35$  & 14.80 & 16.70 & 19.34 & 0.2526 \\
80  & $15.52 \pm 2.06$ & 14.71 & 17.71 & 24.67 & 0.1941 \\
100 & $15.86 \pm 2.40 $ & 15.46 & 18.14 & 24.81 & 0.1587 \\
\bottomrule
\end{tabular}
\caption{Representative compute latency (ms/chunk) for streaming inference at different chunk sizes under $\text{sr}=48$ kHz and hop=320. We report p50/p90/p99 and RTF as $L_{\text{comp}}/T_{\text{chunk}}$.}
\label{tab:latency}
\end{table}

\section{Details of Experiments}
\label{app:experiment_details}
\subsection{Subjective evaluation}
\label{app:subjective_evaluation}
The subjective evaluation is conducted in a controlled acoustic environment featuring sound-attenuated conditions, 
precisely calibrated playback systems, and frequency-equalized headphones to ensure consistency across listening sessions. 
A total of 200 audio segments are randomly sampled from the test dataset for evaluation purposes. 
Our subjective protocol follows the listening-test practice commonly used in long-form speech and song generation research~\cite{li2026swanvoice, zhang2026swantale, zhang2025versatile, pan2025syntheticsingers}.
We recruit 29 participants to provide perceptual ratings across two dimensions: audio quality and spatial perception, 
using a 5-point Likert scale ranging from 1 (Poor) to 5 (Excellent).

For audio quality assessment, we use the Mean Opinion Score for Quality (MOS-Q),
where participants listen with headphones and evaluate the clarity and naturalness of the synthesized audio.
For spatial perception assessment, we adopt the Mean Opinion Score for Position (MOS-P), 
where participants judge the authenticity of spatial attributes, 
including the correspondence between the perceived sound source localization (direction and distance) and the textual prompt specifications. 

\begin{figure*}[p]  
  \centering
  \includegraphics[width=0.9\textwidth]{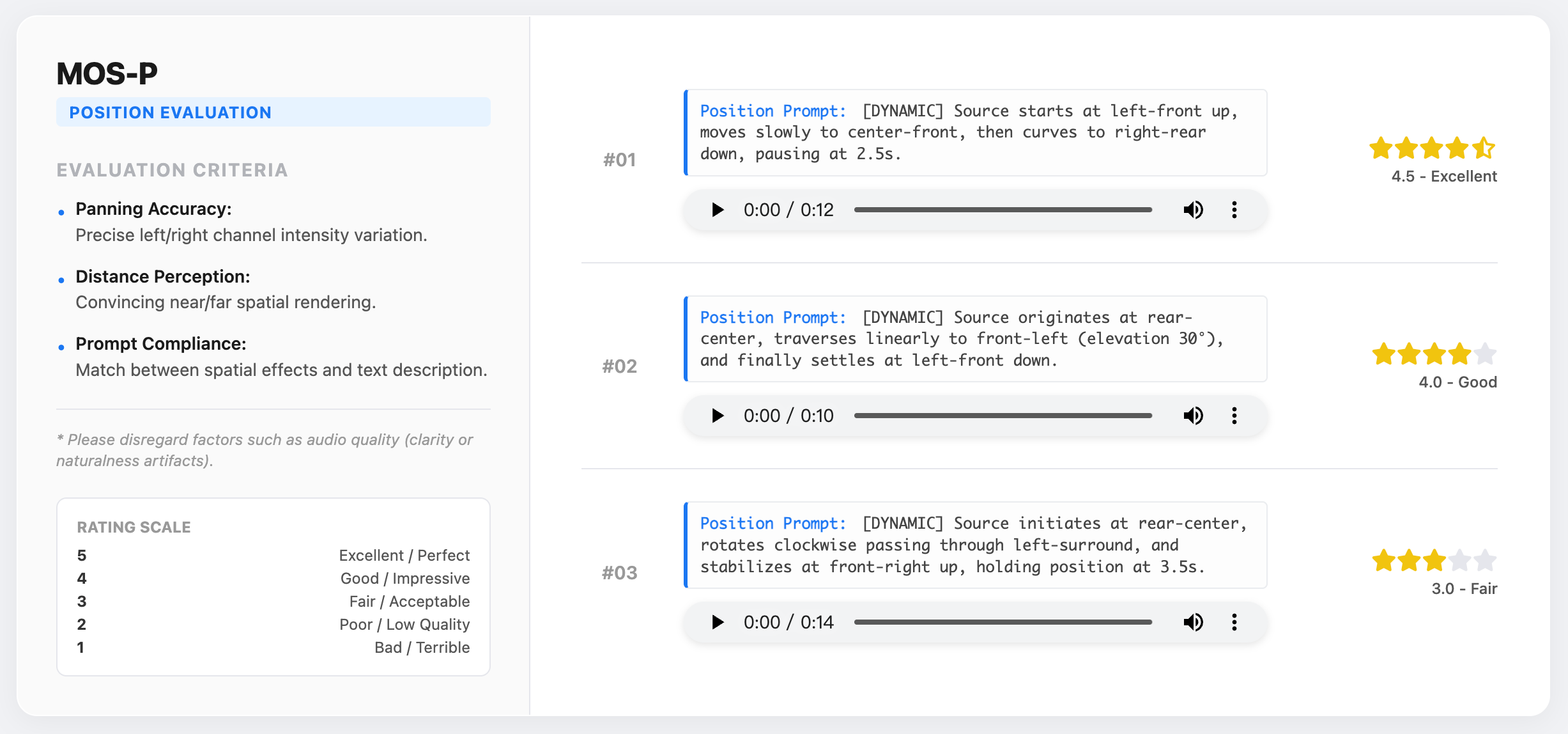}
  \caption{Screenshot of the MOS-P test interface.}
  \label{fig:mosp}
\end{figure*}

\begin{figure*}[p] 
  \centering
  \includegraphics[width=0.9\textwidth]{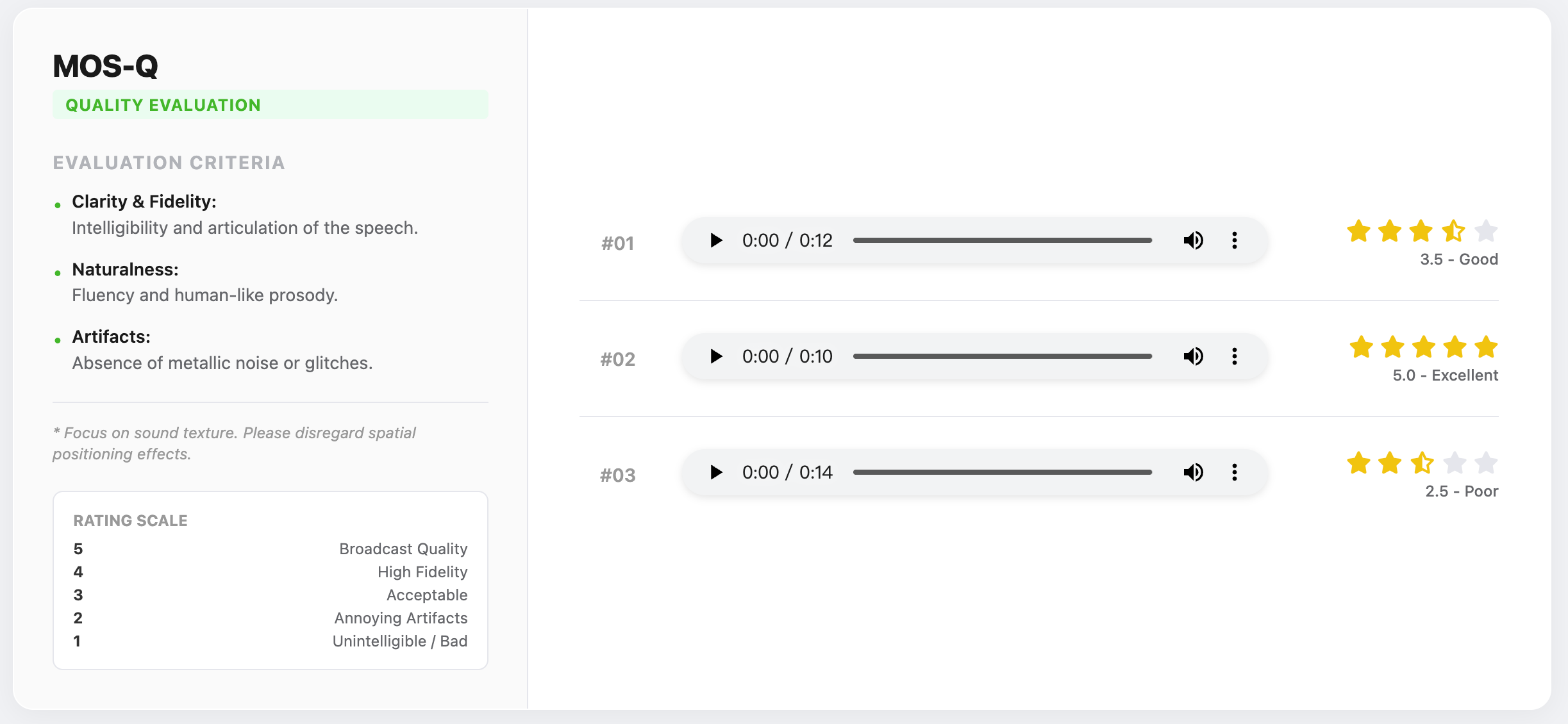}
  \caption{Screenshot of the MOS-Q test interface.}
  \label{fig:mosq}
\end{figure*}

All participants receive appropriate compensation at an hourly rate of \$20, yielding a total experimental cost of approximately \$1500. 
Prior to participation, subjects are informed that their assessments will be used exclusively for academic research purposes.
Detailed instructions for the audio evaluation protocol are shown in Figures~\ref{fig:mosp} and~\ref{fig:mosq}.

\subsection{MUSHRA-style Expert Evaluation}
\label{app:mushra}
To complement the MOS study with non-expert listeners, we further conduct a MUSHRA-style expert test following the ITU-R BS.1534 methodology~\cite{schoeffler2015towards}, focused on spatial perception rather than audio quality alone.
The test involves 9 spatial audio experts and 120 dynamic spatial audio generation results.
The evaluated systems are our method and 5 baselines (Vocos, WaveFM, DiffWave, PriorGrad, and FastDiff).
Each trial contains an explicit reference, a hidden reference, and a 3.5 kHz low-pass / dual-mono anchor.
Participants rate each stimulus on a 0 to 100 scale.
We report quality scores (MUSHRA-Q) and spatial perception scores (MUSHRA-P), where MUSHRA-P covers direction, distance, externalization, and image stability.

Table~\ref{tab:mushra} shows the results with 95\% confidence intervals.
CSAVocoder achieves the highest MUSHRA-P (88.0) and overall score (85.5).
Its MUSHRA-Q is slightly below Vocos and PriorGrad, but its spatial perception score exceeds all baselines by a large margin, which is consistent with the objective spatial metrics.
The hidden reference scores close to 100 and the anchor close to the lower bound, confirming that the protocol is reliable.

\begin{table}[t]
\centering
\footnotesize
\setlength{\tabcolsep}{2pt}
\begin{tabular}{lccc}
\toprule
\textbf{System} & \textbf{Q} ($\uparrow$) & \textbf{P} ($\uparrow$) & \textbf{Overall} ($\uparrow$) \\
\midrule
Hidden ref. & 96.0 {\scriptsize[95.2, 96.8]} & 96.5 {\scriptsize[95.6, 97.4]} & 96.3 {\scriptsize[95.6, 96.9]} \\
CSAVocoder       & 83.0 {\scriptsize[81.8, 84.2]} & \textbf{88.0} {\scriptsize[86.6, 89.3]} & \textbf{85.5} {\scriptsize[84.5, 86.6]} \\
Vocos            & \textbf{86.5} {\scriptsize[85.5, 87.5]} & 70.0 {\scriptsize[68.1, 71.9]} & 78.3 {\scriptsize[76.8, 79.7]} \\
WaveFM           & 84.7 {\scriptsize[83.7, 85.7]} & 72.0 {\scriptsize[70.1, 73.9]} & 78.4 {\scriptsize[77.0, 79.7]} \\
DiffWave         & 78.8 {\scriptsize[77.7, 80.0]} & 58.0 {\scriptsize[56.1, 60.0]} & 68.4 {\scriptsize[66.9, 70.0]} \\
PriorGrad        & 85.2 {\scriptsize[83.9, 86.4]} & 60.0 {\scriptsize[58.1, 61.9]} & 72.6 {\scriptsize[71.0, 74.2]} \\
FastDiff         & 71.6 {\scriptsize[70.3, 72.9]} & 55.0 {\scriptsize[53.1, 56.9]} & 63.3 {\scriptsize[61.7, 64.9]} \\
Anchor           & 24.5 {\scriptsize[22.9, 25.9]} & 19.5 {\scriptsize[17.8, 21.2]} & 22.0 {\scriptsize[20.4, 23.6]} \\
\bottomrule
\end{tabular}
\caption{MUSHRA-style expert evaluation. Q: quality, P: spatial perception; 95\% CIs.}
\label{tab:mushra}
\end{table}

\subsection{Objective evaluation}
To ensure the reproducibility of our experiments, we use standard open-source implementations for objective evaluation. The specific configurations and libraries are detailed below:

\vspace{0.5em}
\noindent \textbf{MRSTFT}: We use the Multi-Resolution Short-Time Fourier Transform (MRSTFT) implementation from Auraloss~\cite{steinmetz2020auraloss}. The metric is computed as the sum of spectral convergence and log-magnitude distance across multiple window sizes.
\\
\url{https://github.com/csteinmetz1/auraloss}

\vspace{0.5em}
\noindent \textbf{PESQ}: Perceptual Evaluation of Speech Quality (PESQ) is evaluated using the Wideband mode (ITU-T P.862.2). Since our model generates 48 kHz audio, we downsample both the reference and synthesized signals to 16 kHz solely for this measurement using the \texttt{python-pesq} wrapper.
\\
\url{https://github.com/ludlows/python-pesq}

\vspace{0.5em}
\noindent \textbf{MCD}: We compute the Mel-Cepstral Distortion (MCD) to measure the spectral envelope difference. We use the \texttt{mel-cepstral-distance} library with Dynamic Time Warping (DTW) enabled to align the sequences before calculation.
\\
\url{https://github.com/MattShannon/mcd}

\vspace{0.5em}
\noindent \textbf{Periodicity}: To evaluate pitch accuracy and harmonic consistency, we calculate the periodicity error using the pre-trained CREPE model provided in the CARGAN repository~\cite{cargan}. The metric represents the root mean squared error between the periodicity vectors of the ground truth and generated audio.
\\
\url{https://github.com/descriptinc/cargan}

\vspace{0.5em}
\noindent \textbf{ANG COS \& DIS COS}: To quantify spatial fidelity, we use the pre-trained Spatial-AST model~\cite{zheng2024bat} to extract high-level spatial representations. We report the metrics as ANG COS (for angular consistency) and DIS COS (for distance consistency), where higher cosine similarity indicates better preservation of perceptible spatial cues.
\\
\url{https://github.com/zszheng147/Spatial-AST}

\vspace{0.5em}
\noindent \textbf{RTF}: Real-Time Factor (RTF) is calculated as the time required to generate the waveform divided by the duration of the audio on a single NVIDIA 4090 GPU.

\section{Licenses and Availability}
We respect the original licenses of all referenced artifacts and do not redistribute them. This work uses publicly available datasets. We do not redistribute any third-party audio content. Users must obtain the original datasets from their respective providers and comply with the original licenses/terms of use. We will release only derived metadata (e.g., file lists, splits, and non-invertible statistics) under CC BY 4.0, subject to the original dataset terms. Our codebase may depend on third-party libraries; these components remain under their respective licenses. Any external assets (e.g., pretrained backbones or evaluation tools) are used in accordance with their original licensing terms.

\section{Use of AI Assistants}
We used an AI-based writing assistant during manuscript preparation solely for language polishing, including grammar checking, spelling correction, and improving clarity and readability of the text. All technical claims, experimental procedures, and interpretations were produced and verified by the authors.

\end{document}